# Directional Manipulation of a Staggered Charge Density Wave and Kondo Resonance in UTe$_2$

Nileema Sharma[1,2][†], Fangjun Cheng[1,2][†], Hyeok Jun Yang[1][†], Matthew Toole[1,2], James McKenzie[1,2], Mitchell M. Bordelon[3], Sean M. Thomas[3], Priscila F. S. Rosa[3], Yi-Ting Hsu[1]*, and Xiaolong Liu[1,2]*

[1]Department of Physics and Astronomy, University of Notre Dame, Notre Dame, IN 46556, USA
[2]Stavropoulos Center for Complex Quantum Matter, University of Notre Dame, Notre Dame, IN 46556, USA
[3]Los Alamos National Laboratory, Los Alamos, New Mexico 87545, USA

* Corresponding authors. yhsu2@nd.edu, xliu33@nd.edu
† These authors contributed equally to this work.

**Abstract**

UTe$_2$ is a rare example of a correlated quantum material in which unconventional density wave orders, Kondo physics, spin-triplet pairing, and reentrant superconductivity coexist within the same electronic system. Its superconducting state develops out of a strongly correlated normal phase. The identification and control of competing or intertwined normal-state orders are thus central to elucidating the electronic landscape from which its superconductivity arises. Here, using scanning tunnelling microscopy (STM) in a vector magnetic field, we uncover a previously unreported staggered charge-density-wave (CDW) in high-quality UTe$_2$ crystals and demonstrate its strong directional response to an external magnetic field: the staggered CDW is completely quenched by a modest 1.7 T field aligned with the quasi-one-dimensional uranium chain direction (*a*-axis), while remaining robust against fields along other crystallographic directions. This pronounced anisotropy is consistent with an orbital-driven mechanism that leads to a field-tuned quantum phase transition. Strikingly and counterintuitively, the same field orientation and strength concomitantly alter the hybridization gap and suppress the 5*f* Kondo resonance. Modelling indicates that this correlated evolution arises from a switch of the dominant hybridization channel from Te 5*p*- U 5*f* to U 6*d*- U 5*f* coupling, suggesting an intimate interplay between CDW and the Kondo effect. Our work establishes an effective tuning knob for the intertwined orders in UTe$_2$ and provides evidence for orbital-selective Kondo hybridization, shedding light on its correlated normal state.

**Main**

Strongly correlated materials often exhibit a complex landscape of intertwined orders in which multiple broken-symmetry phases can emerge and compete under external tuning parameters, such as temperature, doping, pressure, and magnetic field[1,2]. This paradigm is epitomized by the high-temperature cuprate superconductors, where charge/spin orders, nematicity, pseudogap behavior, and superconductivity are intricately entangled across their phase diagrams[3,4], yet the interplay between density waves and *d*-wave superconductivity remains an open question[5]. Recently, the heavy-fermion superconductor UTe$_2$ has emerged as a leading candidate for bulk spin-triplet superconductivity[6], possessing exceptionally large upper critical fields[7], re-entrant and field-reinforced superconducting phases[8], exotic vortex structures[9,10,11], and the existence of antiferromagnetic spin correlations[12]. As in the cuprates, its normal state is highly nontrivial and likely bears an intimate connection to the properties of the superconducting state. Notably, UTe$_2$ simultaneously hosts charge density wave (CDW) order[13,14,15,16] and Kondo lattice physics[17]—an unusual coexistence, as both phenomena typically compete for spectral weight near the Fermi level[18]. Previous work has shown that the CDW in UTe$_2$ exhibits unconventional behavior, including being sensitive to high magnetic fields of ~10 T applied nearly along the [011] direction[13,15,16]. At the same time, theoretical studies suggest that Kondo hybridization in UTe$_2$ is orbital selective, involving U-5*f* states coupled to either Te-5*p* or U-6*d* orbitals[19]; however, experimental evidence supporting this scenario has not yet been established. These observations raise several key questions: How do CDW order and Kondo hybridization interact within the highly anisotropic electronic structure of UTe$_2$? How are these interactions modified by external perturbations such as magnetic fields? And can magnetic-field tuning reveal signatures of orbital-selective Kondo hybridization?

Structurally, UTe$_2$ is composed of quasi-one-dimensional (1D) atomic chains extending along the crystallographic *a*-axis (Fig. 1a, inset). This low-dimensional architecture gives rise to two orthogonally oriented sets of hybridized quasi-1D Fermi surfaces derived predominantly from Te-5*p* and U-6*d* orbitals (Supplementary Fig. 1)[20]. In such quasi-1D systems, both the Pauli and orbital effects play essential roles in governing the directional response of charge order to external magnetic fields[21,22,23,24,25,26]. The Pauli effect, arising from spin splitting, is largely isotropic and tends to suppress CDW orders[21,22]. By contrast, the orbital effect on CDW depends sensitively on the orientation of the magnetic field relative to the 1D chains, owing to the anisotropic electronic dispersion[21,22]. Given the generally competing nature of CDW order and Kondo screening[18,27], together with the pronounced structural anisotropy and orbital-selective Kondo hybridization in UTe$_2$, the influence of a vector magnetic field on these intertwined phenomena is expected to be highly nontrivial, yet completely unexplored.

Here, using spectroscopic-imaging scanning tunneling microscopy (SI-STM) on high-quality UTe$_2$ single crystals with a critical temperature $T_c = 2.1$ K, we discover a new CDW order forming a staggered, brick-wall–like modulation in real space. By applying a 3D vector magnetic field, we uncover an exceptionally strong directional field response of both the staggered CDW and Kondo resonance. While the CDW and Kondo spectral features remain virtually unaltered under magnetic fields up to 8.8 T applied perpendicular to the (011) surface, the staggered CDW undergoes a field-driven quantum phase transition and becomes completely suppressed when a comparatively modest field of 1.7 T is rotated toward the crystallographic *a*-axis in the normal state. Notably, and counter to conventional expectations, the Kondo resonance is simultaneously suppressed as the CDW vanishes. The concomitant suppression of the staggered CDW and the Kondo resonance with *a*-axis magnetic field is consistent with field-induced orbital effects and the existence of orbital-selective Kondo effects in UTe₂. More specifically, an *a*-axis field destabilizes the staggered CDW, which is derived from nesting with quasi-1D U-6*d* bands, and shifts the dominant Kondo hybridization from *p-f* to *d-f* channels.

**Discovery of a new staggered CDW**

Cryogenic cleaving of UTe₂ single crystals exposes the (011) plane as shown in Fig. 1a. Differential conductance imaging [$g(\mathbf{r}, V) \equiv dI/dV$], proportional to the local density of states (LDOS), resolves not only the Te atomic lattice (Fig. 1b), but additional modulation wavevectors as most clearly seen in the Fourier transform image $g(\mathbf{q}, 10 \text{ mV})$ in Fig. 1c acquired at 0.3 K. In addition to the previously reported[13,14,15,16] CDW wavevectors $Q_i$, we discovered a new set of distinct wavevectors $W_i$, where $i = 1,2,3$. These wavevectors are non-dispersive with energy and exhibit contrast inversion across zero bias (Supplementary Fig. 2) – behavior characteristic of CDWs arising from Fermi surface nesting and accompanied by a gap opening at the Fermi level[28,29]. These observations establish $W_i$ as a new set of CDWs in UTe₂ with $W_{1,3} = (\mp \frac{2\pi}{b^*}, 0.13 \frac{2\pi}{a})$, where $a$ and $b^*$ are defined in Fig. 1a inset. By visualizing the thermal melting of both $W_i$ and $Q_i$ CDWs, we determined that the $W_i$ CDWs persist into the normal state of UTe₂ with a critical temperature $T_{CDW}^{W_i} \sim 6$ K (Supplementary Figs. 3). At the bias where the $W_i$ CDW intensity peaks (3.33 mV, Fig. 1d), the real-space pattern forms a staggered, brick-wall–like modulation, with periodic hotspots arranged alternately along neighboring Te chains (Fig. 1e). This structure closely resembles Coulomb coupled 1D CDWs [30] described by the Šaub-Barišić-Friedel model [31], suggesting the quasi-1D nature of the $W_i$ CDW. Consistent with this picture, both $W_i$ and $Q_i$ wavevectors connect segments of the warped quasi-1D U-derived bands[32] and lie on the edges of the surface Brillouin zone (Supplementary Fig. 1), analogous to quasi-1D CDWs in organic conductors[33] from Fermi surface nesting. Furthermore, spatial correlations between the mirror symmetric $W_1$ ($Q_1$) and $W_3$ ($Q_3$) modulations are observed (Supplementary Figs. 4,5). Compared to $W_1$ and $W_3$, the $W_2$ wavevector has a much lower intensity close to background noise (Fig. 1c,f), possibly due to the presence of quenched disorders and/or the formfactor of $W_i$ CDW (Supplementary Fig. 6). Thus, we focus on $W_1$ and $W_3$ in this study. Although $Q_1 \cong W_1 + W_2$, the set of $W_i$ CDW constitutes an independent electronic order because $W_i$ and $Q_i$ CDWs exhibit vastly different energy dependence of their intensities (Fig. 1d), critical temperatures, critical fields (Fig. 3), as well as uncorrelated spatial distributions of topological defects (Supplementary Figs. 4, 5).

**Insensitivity of CDW and Kondo resonance to out-of-plane magnetic fields**

Having established the new $W_i$ CDW order, we explore the effect of an out-of-plane magnetic field $B_\perp$ along the (011) surface normal direction (defined as c*-axis, Fig. 1a inset). Figure 1d compares the energy-dependence of the $W_1$ and $Q_1$ intensities under different $B_\perp$ and field-sweep directions. The nearly overlapping curves suggest their minimal sensitivity to an out-of-plane magnetic field up to 8.8 T. This robustness is better illustrated in Fig. 2a, where the integrated intensities of $W_1$ (at $V = 3.33$ mV) and $Q_1$ (at $V = 10$ mV) in momentum space, corresponding to their respective peak amplitudes, remain essentially unchanged across the entire field range. Real-space $g(r, V)$ and the extracted CDW amplitude/phase images of the same field of views (FOVs) under different $B_\perp$ likewise show no obvious evolution with $B_\perp$ (Supplementary Figs. 4,5), further demonstrating the robustness and reproducibility of our STM imaging. The insensitivity of the CDWs to $B_\perp$ is consistent with the negligible theoretical reduction $\Delta T_{\text{CDW}} = -\frac{\gamma}{4}\left(\frac{\mu_B B}{k T_{\text{CDW}}(0)}\right)^2 T_{\text{CDW}}(0)$ due to the Pauli effect[21,22]. Here, $\gamma$ is of order 1. For $T_{\text{CDW}}^{W_i} \approx 6$ K and $T_{\text{CDW}}^{Q_i} \approx 10$ K (Supplementary Fig. 3), the Pauli-limiting fields are approximately $B_P^{W_i} \approx 18$ T and $B_P^{Q_i} \approx 30$ T, respectively. Indeed, the $Q_i$ CDW has recently been detected[34] even at $B_\perp = 20$ T.

We next examine the influence of $B_\perp$ on the Kondo effect in UTe$_2$ spectroscopically. Figure 2b presents a set of spatially averaged differential conductance spectra measured at 0.3 K under different $B_\perp$, exhibiting virtually identical spectra with a broad spectral bump near –4 mV (red arrow) and a dip at zero bias. The bump is consistent with previously reported[17] Kondo resonance signature in UTe$_2$. After subtracting a linear background (see Methods), this spectral feature resolves into two peaks that are ~2 mV apart as indicated by the blue and green arrows in Fig. 2c. These two peaks are characteristic of a Kondo lattice[35,36], arising from hybridization between a light conduction band (grey) and a heavy *f*-electron band (blue) (inset of Fig. 2c), which produces asymmetric coherence peaks separated by a hybridization gap $\Delta_h$. The extracted peak heights ($H_{1,2}$) and hybridization gap exhibit negligible variation with $B_\perp$ up to 8.8 T (Fig. 2d), indicating that Kondo coherence remains robust, consistent with behaviors observed in other heavy-fermion systems under comparable field strengths[37,38].

We further examine whether the staggered CDW is coupled to or competes with spin-triplet superconductivity in UTe$_2$. To this end, we image vortex lattices (VLs) generated under identical magnetic fields but separate field cooling procedures. Because individual vortices – where superconductivity is locally suppressed – nucleate at different spatial locations in repeated field-cooling cycles, this approach enables us to test whether the CDW responds to local suppression of superconductivity. As shown in Figs. 2e,f, vortices in UTe$_2$ exhibit a characteristic doublet structure[9,10,11], and the two triangular VLs created at $B_\perp = 8.8$ T are clearly displaced from each other in the zero-bias conductance images [$g(r, 0\text{ V})$ and $g'(r, 0\text{ V})$] of the same field of view. The images are registered to each other via a multi-image registration procedure (see Methods).

However, the corresponding $g(r, 3.33\text{ mV})$ images, where the staggered $W_i$ CDW is most pronounced, show virtually no changes (Figs. 2g,h). This is further confirmed by examining the difference maps $[\delta g(r,V) \equiv g(r,V) - g'(r,V)]$ in Figs. 2i,j: while $\delta g(r, 0\text{ V})$ clearly reflects the displaced VL lattice, the $\delta g(r, 3.33\text{ mV})$ image is essentially noise. Moreover, phase-resolved analysis using a 2D lock-in algorithm (see Methods) shows that the spatial phase $\Phi_{W_1}(r)$ of the $W_1$ CDW and its associated topological defects in the form vortices and anti-vortices[39] (indicated by the white and black dots, respectively) remain unchanged (Figs. 2k,l). This further confirms that the $W_i$ CDW is insensitive to local suppression of superconductivity and thus couples only weakly to uniform superconductivity.

**Simultaneous manipulation of CDW and Kondo resonance with an in-plane magnetic field**

Next, we investigate the responses of CDW and Kondo effect to a vector magnetic field $\boldsymbol{B}_\parallel$ applied in the (011) plane with an angle $\theta$ from the $a$-axis. Figure 3a displays a series of $g(\boldsymbol{q}, 3.33\text{ mV})$ images that capture the evolution of $W_i$ CDW as $\boldsymbol{B}_\parallel$ is rotated within the plane at 2.5 K [full sets of $g(r, 3.33\text{ mV})$, $g(r, 10\text{ mV})$ and their Fourier transforms for $\theta$ varying between 0 and $2\pi$ are given in Supplementary Figs. 7,8]. Remarkably, although the field strength $B_\parallel = 1.7$ T is moderate, a complete (strong) suppression of the staggered $W_i$ ($Q_i$) CDW is observed when $\boldsymbol{B}_\parallel$ is parallel to the $a$-axis ($\theta = 0$) of UTe$_2$, whereas virtually no effects are present when $\boldsymbol{B}_\parallel$ is perpendicular ($\theta = \pi/2$) to the $a$-axis (defined as the $b^*$-axis, Fig. 1a inset; Supplementary Fig. 9). Such highly directional response of CDW to $\boldsymbol{B}_\parallel$ is reproduced at 0.3 K (Supplementary Fig. 10). Spectroscopically, this suppression of CDW manifests as a pronounced angular modulation of the low-bias conductance (Fig. 3b). At $\theta = 0$ and $\pi$, where the field is parallel to the uranium chains and the CDW is maximally suppressed, the zero-bias conductance increases substantially. Since UTe$_2$ is in the normal state at 2.5 K, such changes are unrelated to superconductivity, but consistent with the closing of CDW gaps around the Fermi level[40]. Real-space maps (Fig. 3c) show that the brick-wall modulation amplitude from $W_i$ CDW continuously diminishes as the field rotates toward the $a$-axis, yielding a two-fold symmetric suppression of CDW intensity (Figs. 3f,g). CDW amplitude $A_{W_1}(r)$ and phase $\Phi_{W_1}(r)$ analysis (Figs. 3d,e) reveals that domains of high CDW amplitude shrink in size without a proliferation of topological defects, indicating that the quantum melting of the CDW is driven by a reduction of the electron–hole pairing amplitude rather than by a loss of phase stiffness[41,42]. This behavior is consistent with the suppression of $T_{\text{CDW}}^{W_i}$ to zero at a critical in-plane field $B_c$ along the $a$-axis of UTe$_2$.

Anisotropy of the Pauli effect cannot fully account for such observations. Although the magnetic susceptibility $\chi$ of UTe$_2$ along the magnetic easy axis ($a$-axis) is about 5 times larger than those along perpendicular directions at temperatures approaching zero[43], the corresponding Pauli paramagnetic limit $B_P = 2\Delta_0/\sqrt{2}g\mu_B \propto 1/\sqrt{\chi}$ (where $\Delta_0$ is the CDW gap, $g$ is the $g$-factor of electrons) is reduced only by a factor of 0.45 along the $a$-axis[44], insufficient to explain why $B_\parallel = 1.7$ T along the $a$-axis fully quenches the CDW, whereas $B_\perp = 8.8$ T along the $c^*$-axis produces no discernable effects. Another possible mechanism is a magnetic field induced Lifshitz transition. While such Fermi surface reconstructions have been observed[45] at the metamagnetic transitions of UTe$_2$ at ~35 T, low-field transport anomalies[46] appear only above fields of 5.6 T along the *a*-axis, making Lifshitz transition unlikely to account for the observed CDW suppression with $B_\parallel = 1.7$ T. By contrast, it has been established both theoretically[21,23,24] and experimentally[22,25,26] that when an external magnetic field is applied perpendicular or parallel to the chains of a quasi-1D system, the CDW therein will be stabilized or suppressed, respectively, resulting from confined orbital motion of charges. In UTe$_2$, the 1D uranium chains run along the *a*-axis while both the b*- and c*- axes are transverse to the chains. It is therefore plausible to suspect that the orbital effect of electrons is responsible for the observed highly anisotropic response of CDW to vector magnetic fields. To better visualize this behavior, we map the integrated intensities of $W_i$ and $Q_i$ CDWs in the phase space of $B_\parallel$, $\theta$, and $T$, as shown in Figs. 3h,i. The phase boundaries defining the CDW states are determined by fitting the experimental integrated intensity of CDW peaks (details given in Methods), and the CDW occupy the phase space with $T \to 0$, $B_\parallel \to 0$, and $\theta \to \pi/2$ as shown in Figs. 3h,i. For $W_1$ CDW, the fitting yields a zero-temperature critical field of $B_c^{W_1}(T=0) \approx 1.7$ T along the *a*-axis, consistent with experimental observations (Fig. 2) and the fact that $W_i$ is not fully suppressed at $B_\parallel = 1.5$ T along the *a*-axis at 0.3 K (Supplementary Fig. 10). These results point to a putative quantum critical point, beyond which the $W_i$ CDW becomes quantum disordered at zero temperature.

In light of the potential interplay between CDW orders and Kondo physics, it is worthwhile to examine how $\boldsymbol{B}_\parallel$ simultaneously influences the Kondo resonance. Figure 4a presents a series of spatially averaged d$I$/d$V$ spectra acquired at 0.3 K (for optimal energy resolution) on a different UTe$_2$ crystal under an in-plane magnetic field of $B_\parallel = 1.95$ T, while the angle $\theta$ between $\boldsymbol{B}_\parallel$ and the *a*-axis is varied from 0 to $2\pi$. Consistent with the behavior observed at 2.5 K (Fig. 3b), the zero-bias conductance increases when the CDW is suppressed at $\theta = 0$ and $\pi$, and decreases when it is restored at $\theta = \pi/2$ and $3\pi/2$ as indicated by the orange curve. Meanwhile, the Kondo resonance feature around –4 mV exhibits an opposite angular modulation as indicated by the blue curve. After subtracting a linear background (Fig. 4b), the modulation of the Kondo resonance becomes more evident in the extracted peak heights ($H_{1,2}$) and the hybridization gap ($\Delta_\text{h}$). The polar plots in Figs. 4c,d demonstrate an in-phase modulation between $H_1$ and $H_2$, and an out-of-phase modulation between $H_{1,2}$ and $\Delta_\text{h}$ as $\theta$ is varied from 0 to $2\pi$. Notably, the relative amplitudes of $H_1$ and $H_2$ invert between $\theta = 0$ (or $\pi$) and $\theta = \pi/2$ (or $3\pi/2$).

Analogous to the response of CDW to $\boldsymbol{B}_\parallel$, the anisotropic Pauli effect alone is unlikely to account for this pronounced directional modulation by $\boldsymbol{B}_\parallel$, particularly given the insensitivity of the Kondo resonance to $B_\perp$ up to 8.8 T. The distinct angular evolution of $H_{1,2}$ and $\Delta_\text{h}$ further points to an additional mechanism. Because both Te-5$p$ and U-6$d$ itinerant electrons can participate in Kondo screening[19] that competes with CDW formation[18,27], and because both the $W_i$ and $Q_i$

CDWs likely originate from nesting of the U-$6d$ bands (Supplementary Fig. 1), the formation of CDW weakens the overall Kondo screening by U-$6d$-electrons. In contrast, the itinerant $p$-electrons provide a uniform charge background irrespective of the presence of CDWs. It is therefore reasonable to infer that, when CDWs are present, the dominant Kondo hybridization occurs in the $p$–$f$ channel. When the CDWs are suppressed, the proximity between U-$6d$ and U-$5f$ orbitals renders an effective shift in the dominant hybridization channel to $d$–$f$, as illustrated schematically in Figs. 4e,f. With physically reasonable parameters specific to UTe$_2$, we construct a minimal theoretical model incorporating this hybridization reconfiguration proposal (see details in Methods). The model qualitatively reproduces the key experimental signatures in calculated $dI/dV$ spectra (Fig. 4g), including the modulation of $H_{1,2}$ and $\Delta_h$, as well as the inversion in the relative peak heights of $H_1$ and $H_2$ between the CDW and CDW-suppressed states. Although field-induced modifications of exchange coupling likely also contribute[47], they are not a necessary ingredient in our model to reproduce the experimental observations, and a more detailed material-specific microscopic treatment for UTe$_2$ is required to elucidate its role. Therefore, the simultaneous tuning of CDW and Kondo resonance provides direct evidence supporting the orbital-selectivity of the Kondo effect in UTe$_2$.

**Conclusion**

In this work, we discovered a new set of $W_i$ CDW with a staggered real-space structure in high-quality UTe$_2$ crystals. We demonstrate that both the staggered CDW and Kondo coherence are insensitive to magnetic fields applied perpendicular to the quasi-1D chains, yet are concomitantly suppressed by a modest field aligned along the $a$-axis, revealing a putative CDW quantum critical point at $B_\parallel \approx 1.7$ T. Such pronounced vector-field tunning of CDW and Kondo resonance cannot be accounted for by the Pauli effects alone and instead point to an orbital-driven mechanism and a modification of the dominant Kondo hybridization channels. These observations provide direct evidence for orbital-selective Kondo hybridization and intertwined charge order and Kondo screening in UTe$_2$. Because the $W_i$ CDW persist in the superconducting state of UTe$_2$, it necessarily induces a set of composite spin-triplet pair density waves (PDWs) with the same wavevectors even if the CDWs do not couple strongly to superconductivity (Fig. 2). Our results therefore imply a concomitant magnetic field tuning of such composite PDWs, which deserves future investigation. Overall, our work demonstrates that directional magnetic fields are a powerful tuning knob for controlling emergent order and probing the couplings between them. Our findings also introduce charge order and hybridization switching as key ingredients of the normal state phase diagram of UTe$_2$.

**Methods**

1. Crystal synthesis and STM measurements

As in our previous study[9], high quality UTe$_2$ crystals with $T_c = 2.1$ K are grown using a molten salt flux method and are polished along the (011) direction for mechanical cleaving. The UTe$_2$ crystals are cleaved at cryogenic temperature ($\sim 10$ K) under ultra-high vacuum ($<$ $1 \times 10^{-10}$ Torr) conditions and immediately transferred into the scan head of a Unisoku USM1300J microscope. Mechanically cut Nb wires are used as STM tips after outgassing in ultra-high vacuum conditions. As prepared tips are superconducting under zero magnetic field, but become normal under a magnetic field of 1 T (Supplementary Fig. 11). Therefore, for all data taken under non-zero field in this study, the tip is metallic to eliminate complexities associated with tip superconductivity. All STM measurements are performed using SPECS Nanonis electronics with a built-in lock-in amplifier. MATLAB and Gwyddion software are used for processing the acquired data.

2. Image correction and registration with sub-atomic precision

Because of the ubiquitous thermal drifts and artifacts associated with the piezoelectric scanner in an STM, images taken in nominally identical FOVs but from separate measurements need to be precisely registered with each other with sub-atomic precision to enable rigorous extraction and comparisons of CDW intensity and spatially resolved modulation amplitude, phase, and topological defects. This is achieved in three steps: firstly, the raw experimental images are up-sampled by a factor of four. Secondly, the Lawler-Fujita (LF) algorithm[48] and shear transformation are applied to restore a periodic atomic lattice from experimental images that contain distortions due to thermal and piezoelectric drifts. Lastly, affine transformations are applied to register multiple images to the same FOV with sub-atomic precision.

A perfectly periodic lattice $\tilde{T}(\tilde{r})$ with Bragg wavevectors $\boldsymbol{B}_i = (B_i^x, B_i^y)$, where $i = 1, 2, 3$, can be written as

$$\tilde{T}(\tilde{r}) = \sum_{i=1}^{3} T_i \cos(\boldsymbol{B}_i \cdot \tilde{r} + \bar{\varphi}_i)$$

Here, $\bar{\varphi}_i$ represents the constant spatial phase shift along $\boldsymbol{B}_i$. Experimentally, the STM topographic image $T(\boldsymbol{r})$ suffers from a slowly varying $\boldsymbol{r}$-dependent spatial phase shift $\varphi_i(\boldsymbol{r})$, leading to

$$T(\boldsymbol{r}) = \sum_{i=1}^{3} T_i \cos(\boldsymbol{B}_i \cdot \boldsymbol{r} + \varphi_i)$$

To extract $\varphi_i(\boldsymbol{r})$, experimental topographic image $T(\boldsymbol{r})$ is multiplied by a reference wave $e^{i\boldsymbol{B}_i \cdot \boldsymbol{r}}$ and integrated over a Gaussian filter with width $\sigma = 1.64$ nm (i.e., a few lattice constants) to obtain the complex-valued $A_{\boldsymbol{B}_i}(\boldsymbol{r})$

$$A_{\boldsymbol{B}_i}(\boldsymbol{r}) = \frac{1}{\sqrt{2\pi}\sigma} \int d\boldsymbol{R}\, T(\boldsymbol{R}) e^{i\boldsymbol{B}_i \cdot \boldsymbol{R}}\, e^{-\frac{(\boldsymbol{r}-\boldsymbol{R})^2}{2\sigma^2}}$$

Such a convolution integral can be conveniently calculated by using the convolution theorem

$$A_{\boldsymbol{B}_i}(\boldsymbol{r}) = F^{-1}[\tilde{A}_{\boldsymbol{B}_i}(\boldsymbol{q})] = F^{-1}\left[F(T(\boldsymbol{r})e^{i\boldsymbol{B}_i \cdot \boldsymbol{r}}) \frac{1}{\sqrt{2\pi}\sigma_q} e^{-\frac{q^2}{2\sigma_q^2}}\right]$$

where $F$ and $F^{-1}$ denote Fourier transform and inverse Fourier transform, respectively, and $\sigma = 1/\sigma_q$. Then we can obtain

$$\varphi_i(\boldsymbol{r}) = \tan^{-1}\left\{\frac{\text{Im}[A_{B_i}(\boldsymbol{r})]}{\text{Re}[A_{B_i}(\boldsymbol{r})]}\right\}$$

Because the relationship between the distorted and the perfect Bragg lattice for each $\boldsymbol{B}_i$ is

$$\boldsymbol{B}_i \cdot \boldsymbol{r} + \varphi_i(\boldsymbol{r}) = \boldsymbol{B}_i \cdot \tilde{\boldsymbol{r}} + \bar{\varphi}_i$$

, we can define a displacement field $\boldsymbol{u}(\boldsymbol{r}) = \boldsymbol{r} - \tilde{\boldsymbol{r}}$ and write it in the matrix form

$$B\boldsymbol{u}(\boldsymbol{r}) = \bar{\varphi} - \varphi(\boldsymbol{r})$$

where $B = (\boldsymbol{B}_1 \ \boldsymbol{B}_2 \ \boldsymbol{B}_3)^{\text{T}}$ is a $3 \times 2$ matrix that is left-invertible, $\boldsymbol{u}(\boldsymbol{r}) = (u_1(\boldsymbol{r}) \ u_2(\boldsymbol{r}) \ u_3(\boldsymbol{r}))^{\text{T}}$ and $\bar{\varphi} - \varphi(\boldsymbol{r}) = (\bar{\varphi}_1 - \varphi_1(\boldsymbol{r}) \ \bar{\varphi}_2 - \varphi_2(\boldsymbol{r}) \ \bar{\varphi}_3 - \varphi_3(\boldsymbol{r}))^{\text{T}}$ are two column vectors. Therefore, we can obtain

$$\boldsymbol{u}(\boldsymbol{r}) = B^{-1}(\bar{\varphi} - \varphi(\boldsymbol{r}))$$

The choice of $\bar{\varphi}_i$ is arbitrary. A nearly perfectly periodic image can be obtained as $T(\boldsymbol{r} + \boldsymbol{u}(\boldsymbol{r}))$. Such LF-corrected image is then sheared, which can be represented by a transformation matrix $S = \begin{bmatrix} 1 & p \\ 0 & 1 \end{bmatrix}$, to obtain a periodic and symmetric lattice.

Lastly, to register multiple images to the exact same FOV, we first define three reference points in each image (e.g., three atomic defects). Using one LF- and shear-corrected image as the reference, an affine transformation matrix

$$\begin{bmatrix} x' \\ y' \\ 1 \end{bmatrix} = \begin{bmatrix} a & b & e \\ c & d & f \\ 0 & 0 & 1 \end{bmatrix} \begin{bmatrix} x \\ y \\ 1 \end{bmatrix}$$

is determined by requiring the 3 reference points in an image to be registered with the ones in the reference image after the affine transformation. Then, the same affine transformation is applied to the entire image.

### 3. CDW phase and amplitude extraction

A periodic modulation of the differential conductance due to CDW with wavevector $\boldsymbol{Q}$ can be expressed as

$$g(\boldsymbol{r}) = A_{\boldsymbol{Q}}(\boldsymbol{r}) \cos\left(\boldsymbol{Q} \cdot \boldsymbol{r} + \Phi_{\boldsymbol{Q}}(\boldsymbol{r})\right) + g_0$$

with amplitude of modulation $A_{\boldsymbol{Q}}$, spatially dependent phase $\Phi_{\boldsymbol{Q}}$, and a constant background $g_0$. To extract $A_{\boldsymbol{Q}}$ and $\Phi_{\boldsymbol{Q}}$, we use a computational 2D-lockin technique similar to that applied in the LF algorithm. Firstly, $g(\boldsymbol{r})$ is multiplied by a reference signal $e^{i\boldsymbol{Q}\cdot\boldsymbol{r}}$ with identical periodicity with the CDW. Then it is integrated over the entire FOV to obtain the complex valued $g_{\boldsymbol{Q}}(\boldsymbol{r})$

$$g_{\boldsymbol{Q}}(\boldsymbol{r}) = \frac{1}{\sqrt{2\pi}\sigma} \int g(\boldsymbol{R}) e^{-\frac{|\boldsymbol{r}-\boldsymbol{R}|^2}{2\sigma^2}} e^{-i\boldsymbol{Q}\cdot\boldsymbol{R}} d\boldsymbol{R}$$

, which similarly is calculated in the momentum space using the convolution theorem. Here, $\sigma$ is the radius of a Gaussian filter in real space that has to be larger than $2\pi/Q$ (i.e., larger than the period) but small enough such that the equivalent $\boldsymbol{q}$-space filter covers the peak at $\boldsymbol{Q}$ in the FT to avoid artificial truncation of the signal. Considering the length scale of the CDWs, we used Gaussian widths $\sigma_{Q_i} = 1.80$ nm and $\sigma_{W_i} = 2.45$ nm for the $Q_i$ and $W_i$ CDW, respectively. The phase $\Phi_{\boldsymbol{Q}}(\boldsymbol{r})$ is therefore obtained as

$$\Phi_{\boldsymbol{Q}}(\boldsymbol{r}) = \tan^{-1}\left\{\frac{\text{Im}[g_{\boldsymbol{Q}}(\boldsymbol{r})]}{\text{Re}[g_{\boldsymbol{Q}}(\boldsymbol{r})]}\right\}$$

The CDW modulation amplitude $A_{\boldsymbol{Q}}(\boldsymbol{r})$ at location $\boldsymbol{r}$ is thus given by the amplitude $|g_{\boldsymbol{Q}}(\boldsymbol{r})|$. If $\Phi_{\boldsymbol{Q}}(\boldsymbol{r})$ is outside $[0, 2\pi]$, it is brought into this range by $\Phi_{\boldsymbol{Q}}(\boldsymbol{r}) \rightarrow mod[\Phi_{\boldsymbol{Q}}(\boldsymbol{r}), 2\pi]$.

4. CDW intensity extraction

FTs of registered conductance maps measured in the same FOV are used for analyzing the intensity of the CDW wavevectors ($Q_i$ and $W_i$). The CDW peak intensity (e.g., Fig. 1d) is determined by locating the maximum intensity of the corresponding wavevector in the FT, and integrate the intensity within a circular region with a 3-pixel (5.58 $nm^{-1}$) radius, which is large enough to include the entire CDW signal but not covering unrelated momentum space features. In temperature dependent measurements (Supplementary Fig. 3), the $W_i$ CDW intensity was found to be within the noise floor at $T = 7$ K. Given the greater background signal at such low-$\boldsymbol{q}$ regions, the intensity obtained under this condition was defined as the background level and was subtracted from all $W_i$ CDW intensity measurements. Since the $Q_i$ CDW have much larger $q$, the background signal is significantly lower and thus we choose not to subtract any background.

5. Plotting the CDW phase boundaries

Since the $a$-axis magnetic field is the non-thermal parameter controlling the CDWs, the $W_i$ and $Q_i$ CDW intensities can be fitted using the power law[49,50]

$$I(B_p, T) = I_0 \left(1 - \frac{T}{T_c(B_p)}\right)^{2\beta(B_p)}$$

where $B_p = B\cos\theta$, $T_c(B_p) = T_c(0)\left(\frac{B_c - B_p}{B_c}\right)$ (ref. 49,50,51), and we approximate $\beta(B_p)$ using a third order polynomial $\beta(B_p) = aB_p^3 + bB_p^2 + cB_p + d$. Here, $B_c$, $T_c(0)$, $a$, $b$, $c$ and $d$ are the fitting parameters. The fitted parameters for $Q_1$ are $B_c = 2.67\ T$, $T_c(0) = 10.18$ K, $\beta(B_p) = -0.707B_p^3 + 2.08B_p^2 - 1.51B_p + 0.465$, and $I_0 = 10.8$. The fitted parameters for $W_1$ are $B_c = 1.73\ T$, $T_c(0) = 6.03\ K$, $\beta(B_p) = 0.0792B_p^3 + 0.266B_p^2 - 0.5B_p + 0.193$, and $I_0 = 3.54$. The phase boundaries shown in Figs. 3h,i are defined by letting $I(B_p, T) = 0$.

6. Extracting experimental Kondo resonance peak features

To quantify the energy positions and heights of the two Kondo-related spectral features, we first smoothen the experimental differential conductance spectra by spline interpolation. Next, a background subtraction is performed to isolate the Kondo resonance features as frequently applied in literature[17,52,53]. The background is modeled as a linear function[53] connecting two reference points at $V = -10 \text{ mV}$ and $V = 0 \text{ V}$. Using the background subtracted spectra (e.g., Fig. 2c), two local maxima are identified, allowing the extraction of their energy separation $\Delta_h$ and the corresponding peak heights $H_1$ and $H_2$.

## 7. Tight-binding model of UTe$_2$

The parameters used to calculate the differential conductance $dI/dV$ are estimated from a tight-binding (TB) model of UTe$_2$ previously proposed[54]. Specifically, we consider the limits where the localized U-5$f$ states are primarily coupled with either U-6$d$ electrons (magnetic field along the $a$-axis; $\theta = 0$) or Te-5$p$ electrons (magnetic field perpendicular to $a$ axis; $\theta = \pi/2$) so that the hybridization between U-6$d$ and Te-5$p$ states is assumed to be negligible. The non-interacting U-6$d$ electrons are described by

$$H_{U-6d}(k_x, k_y, k_z) = \begin{pmatrix} \mu_U - 2t_U \cos k_x - 2t_{ch,U} \cos k_y & -\Delta_U - 2t'_U \cos k_x - 2t'_{ch,U} \cos k_y - 4t_{z,U} e^{-i\frac{k_z}{2}} \cos \frac{k_x}{2} \cos \frac{k_y}{2} \\ -\Delta_U - 2t'_U \cos k_x - 2t'_{ch,U} \cos k_y - 4t_{z,U} e^{i\frac{k_z}{2}} \cos \frac{k_x}{2} \cos \frac{k_y}{2} & \mu_U - 2t_U \cos k_x - 2t_{ch,U} \cos k_y \end{pmatrix}$$

expressed in the basis of two U atoms that form the uranium dimer in the center of the unit cell. Here, $\mu_U$ is the chemical potential, $t_U, t'_U$ are the hopping along the $a$-axis and between U-atoms, $t_{ch,U}, t'_{ch,U}$ are hopping between U chains in the $a$-$b$ plane, $t_{z,U}$ is the hopping along the $c$-axis, $\Delta_U$ is the intradimer overlap respectively. The non-interacting Te-5$p$ electrons is obtained by

$$H_{Te-5p}(k_x, k_y, k_z) = \begin{pmatrix} \mu_{Te} - 2t_{ch,Te} \cos k_x & -\Delta_{Te} - 2t_{Te} e^{-ik_y} - 2t_{z,Te} \cos \frac{k_x}{2} \cos \frac{k_y}{2} \cos \frac{k_z}{2} \\ -\Delta_{Te} - 2t_{Te} e^{ik_y} - 2t_{z,Te} \cos \frac{k_x}{2} \cos \frac{k_y}{2} \cos \frac{k_z}{2} & \mu_{Te} - 2t_{ch,Te} \cos k_x \end{pmatrix}$$

expressed in the basis of the two Te sublattice sites along the Te chain directions. Here, $\mu_{Te}$ is the chemical potential, $\Delta_{Te}$ is the intra-unit cell overlap between two Te atoms. Finally, $t_{Te}$ and $t_{ch,Te}(t_{z,Te})$ are the nearest-neighbor hopping integrals within a Te chain along $b$ direction and between neighboring Te chains along $a$ direction ($c$ direction), respectively. All parameters in $H_{U-6d}$ and $H_{Te-5p}$ are adapted from Ref. 54.

## 8. Kondo resonance model of UTe$_2$

We employ a minimal model $H_{total}$ for the Kondo resonance in UTe$_2$ that encodes our hypothesis of a switch of primary Kondo hybridization channels driven by magnetic field direction: When the external magnetic field is along the $a$-axis ($\theta = 0$), the CDW is suppressed so that U-6$d$ electrons are the main participants due to their vicinity to the localized U-5$f$ states. When the external magnetic field is perpendicular to the $a$-axis ($\theta = \frac{\pi}{2}$), the U-6$d$ electrons contribute less to the Kondo resonance due to the formation of CDW, and the main participating conduction electrons in the Kondo resonance become the Te-5$p$ electrons due to their availability.

In the following, we will first describe the calculation of differential tunneling conductance $dI/dV$ using $H_{\text{total}}$ in two limiting cases: the case where the in-plane magnetic field $\boldsymbol{B}_\parallel \perp a$-axis ($\theta = \pi/2$ in Fig. 4e) and the case where $\boldsymbol{B}_\parallel \parallel a$-axis ($\theta = 0$ in Fig. 4f), where the localized U-5$f$ states dominantly hybridized with Te-5$p$ and U-6$d$ orbitals, respectively. The total Hamiltonian is given by

$$H_{\text{total}} = H_{\text{UTe}_2} + H_{\text{tip}} + H_{\text{UTe}_2-\text{tip}}$$

In $H_{\text{UTe}_2}$, we retain only the conduction electron states relevant to the Kondo resonance: the conduction channel is dominated by U-6$d$ states at $\theta = 0$ and by Te-5$p$ states at $\theta = \pi/2$. The Kondo lattice Hamiltonian is given by

$$H_{\text{UTe}_2} = \sum_{\mathbf{k}\sigma} \epsilon_\mathbf{k} c^\dagger_{\mathbf{k}\sigma} c_{\mathbf{k}\sigma} + J \sum_{\mathbf{i}} \mathbf{S}_\mathbf{i} \cdot \left( \sum_{\alpha\beta} c^\dagger_{\mathbf{i}\alpha} \boldsymbol{\sigma}_{\alpha\beta} c_{\mathbf{i}\beta} \right)$$

where $J$ is the Kondo-coupling strength, $c^\dagger_{\mathbf{k}\sigma}$ creates the conduction electron state with momentum $\mathbf{k}$ and spin $\sigma$, and $\mathbf{S}_\mathbf{i}$ is the spin operator of localized U-5$f$ states at site $\mathbf{i}$. The non-interacting dispersion, $\epsilon_\mathbf{k}$ is obtained by diagonalizing the TB model of UTe$_2$ described in the previous section, i.e. the eigenvalues of U-6$d$ matrix when $\theta = 0$, and of Te-5$p$ orbitals when $\theta = \pi/2$. Within the mean-field framework (e.g. in Ref. 55), the Kondo coupling in $H_{\text{UTe}_2}$ is approximated by introducing the pseudo-fermions, $\mathbf{S}_\mathbf{i} = \sum_{\alpha\beta} f^\dagger_{\mathbf{i}\alpha} \boldsymbol{\sigma}_{\alpha\beta} f_{\mathbf{i}\beta}$,

$$H_{\text{UTe}_2} \approx \sum_{\mathbf{k}\sigma} \epsilon_\mathbf{k} c^\dagger_{\mathbf{k}\sigma} c_{\mathbf{k}\sigma} + v \sum_{\mathbf{i}\sigma} (c^\dagger_{\mathbf{i}\sigma} f_{\mathbf{i}\sigma} + \text{H.c.}) + \epsilon_f \sum_{\mathbf{i}\sigma} f^\dagger_{\mathbf{i}\sigma} f_{\mathbf{i}\sigma}$$

where $v = -J\langle f^\dagger_{\mathbf{i}\sigma} c_{\mathbf{i}\sigma}\rangle$ is the hybridization amplitude, and the energy of localized states $\epsilon_f$ sets the center of Kondo resonance. The Hamiltonian of STM tip electrons $H_{\text{tip}}$ is given by

$$H_{\text{tip}} = \sum_{\mathbf{k}\sigma} \epsilon_\mathbf{k} a^\dagger_{\mathbf{k}\sigma} a_{\mathbf{k}\sigma}$$

Finally, the tunneling between the tip and sample is described by

$$H_{\text{UTe}_2-\text{tip}} = \sum_\sigma a^\dagger_{i_0\sigma} (t_f f_{i_0\sigma} + t_c c_{i_0\sigma}) + \text{H.c.}$$

where $i_0$ is the tip position, and $t_f$ and $t_c$ are the tunneling strengths between the STM tip to the localized and conduction electrons, respectively.

The $dI/dV$ for model $H_{\text{total}}$ is given by[35]

$$\frac{dI}{dV}(\omega = eV - i\Gamma) \propto t_c^2 \frac{1}{D_c} \text{Im} \left[ \left(\frac{\tilde{q}}{v}\right)^2 \frac{D_c}{\omega - \epsilon_f} + \left(1 + \frac{\tilde{q}}{\omega - \epsilon_f}\right)^2 \log\left(\frac{\omega + D_1 - \frac{v^2}{\omega - \epsilon_f}}{\omega - D_2 - \frac{v^2}{\omega - \epsilon_f}}\right) \right]$$

where $\Gamma$ is the disorder broadening, and $\tilde{q} \propto \frac{v t_f}{J t_c}$ captures the ratio between the tunneling channels and determines the asymmetry of Fano line shape. Here, we treat $\tilde{q}$ as an effective parameter constrained by physically reasonable values for each field direction $\theta$. Moreover, $-D_1$ and $D_2$ denote the lower and top band edges measured from the chemical potential and $D_c = D_1 + D_2$ is the bandwidth whose inverse is proportional to the density of state. In this analytic form, $\tilde{q}, v, t_c$ are treated as control parameters constrained within physically reasonable ranges and determined by fitting to the experimental data. We then obtain the parameters $D_1$ and $D_2$ for U-6$d$ and Te-5$p$ electrons from the tight-binding models $H_{U-6d}$ and $H_{Te-5p}$ given in the previous section, respectively.

The dependence of $dI/dV(\theta)$ on the external magnetic field direction $\theta$ is controlled by the variable set $(D_1, D_2, \tilde{q}, v, t_c)$. The values of these variables are determined by the conduction channel that participates in Kondo resonance: At $\theta = 0$, we set $(D_1, D_2, \tilde{q}, v, t_c)$ to take the values for U-6$d$ electrons, estimated from $H_{U-6d}$. At $\theta = \frac{\pi}{2}$, we set $(D_1, D_2, \tilde{q}, v, t_c)$ to take the values for Te-5$p$ electrons, estimated from $H_{Te-5p}$. In contrast, $(t_f, \epsilon_f)$ is set by the localized U-5$f$ states and is independent of the field direction. We choose $v_d > v_p$ since the hybridization strength between U-5$f$ and U-6$d$ states is expected stronger than that between U-5$f$ and Te-5$p$ states due to proximity. Meanwhile, we set the tunneling strengths from the STM tip to the U-6$d$ and Te-5$p$ conduction electrons to be $t_d < t_p$ since the Te-5$p$ electrons are spatially closer to the tip on the (011) surface (Figs. 4e,f). The specific values of TB and fitting parameters used in Fig. 4g at $\theta = 0$ and $\pi/2$ are $t_U = 150, t'_U = 80, t_{ch,U} = 10, t'_{ch,U} = 0, t_{z,U} = -30, \Delta_U = 400, t_{Te} = -1500, t_{ch,Te} = 0, t_{z,Te} = -50, \Delta_{Te} = -1500$, fitted from density functional theory and taken from Ref. 54. The other fitting parameters are taken to be $\mu_U = 200, \mu_{Te} = -2000, \epsilon_f = -3, \tilde{q}_d = 500, \tilde{q}_p = 300, v_d = 40, v_p = 35, t_d = 1.5, t_p = 3.5, \Gamma = 0.7$, where all units are in meV.

For intermediate orientation $0 < \theta < \pi/2$, we retain the same analytical form of $dI/dV$ and calculate the $\theta$-dependence by interpolating the model parameters between these two limits. For example, we parametrize the tunneling strength between the tip and conduction electrons as $t_c(\theta) = t_d \left(1 - \frac{\theta}{\pi/2}\right) + t_p \frac{\theta}{\pi/2}$, where $t_c(\theta = 0) = t_d$ and $t_c(\theta = \pi/2) = t_p$. Other variables $D_1$, $D_2$, $\tilde{q}$, and $v$, are interpolated in a similar way for $0 < \theta < \pi/2$.

**Acknowledgements**
The authors thank D. Morr, H. Pirie, B. Janko, B. Assaf, E. Huang, and Q. Gu for valuable discussions. Experimental STM investigations were supported by the U.S. Department of Energy (DOE), Office of Science, Office of Basic Energy Sciences under Award Numbers DE-SC0025021. Theoretical modeling was supported by the National Science Foundation Grant No. DMR-2238748. N.S. and M.T. acknowledge support from the Notre Dame Materials Science and Engineering Fellowship. UTe$_2$ crystal synthesis at the Los Alamos National Laboratory was supported by the U.S. Department of Energy, Office of Basic Energy Sciences, Division of Materials Science and Engineering under project "Quantum Fluctuations in Narrow-Band

Systems." M.M.B. acknowledges support from the Los Alamos Laboratory Directed Research and Development program.

**Author contributions**

X.L. conceived the project. N.S., M.T., and J.M. performed the measurements. M.M.B, S.M.T., and P.F.S.R. synthesized the crystals. Y.-T.H. and H.J.Y.performed theoretical and numerical modeling. N.S., F.C., M.T., and J.M. performed data analysis. X.L. wrote the manuscript with input from all authors. All authors contributed to data interpretation.

**Data availability**

All data needed to evaluate the conclusions in the paper are present in the paper and/or the Supplementary Information. Additional data related to this paper may be requested from the authors.

**Code availability**

Gwyddion and MATLAB are used for data analysis. The code is available from the corresponding authors on reasonable request.

**Competing financial interests**

The authors declare no competing financial interests.

**Figures**

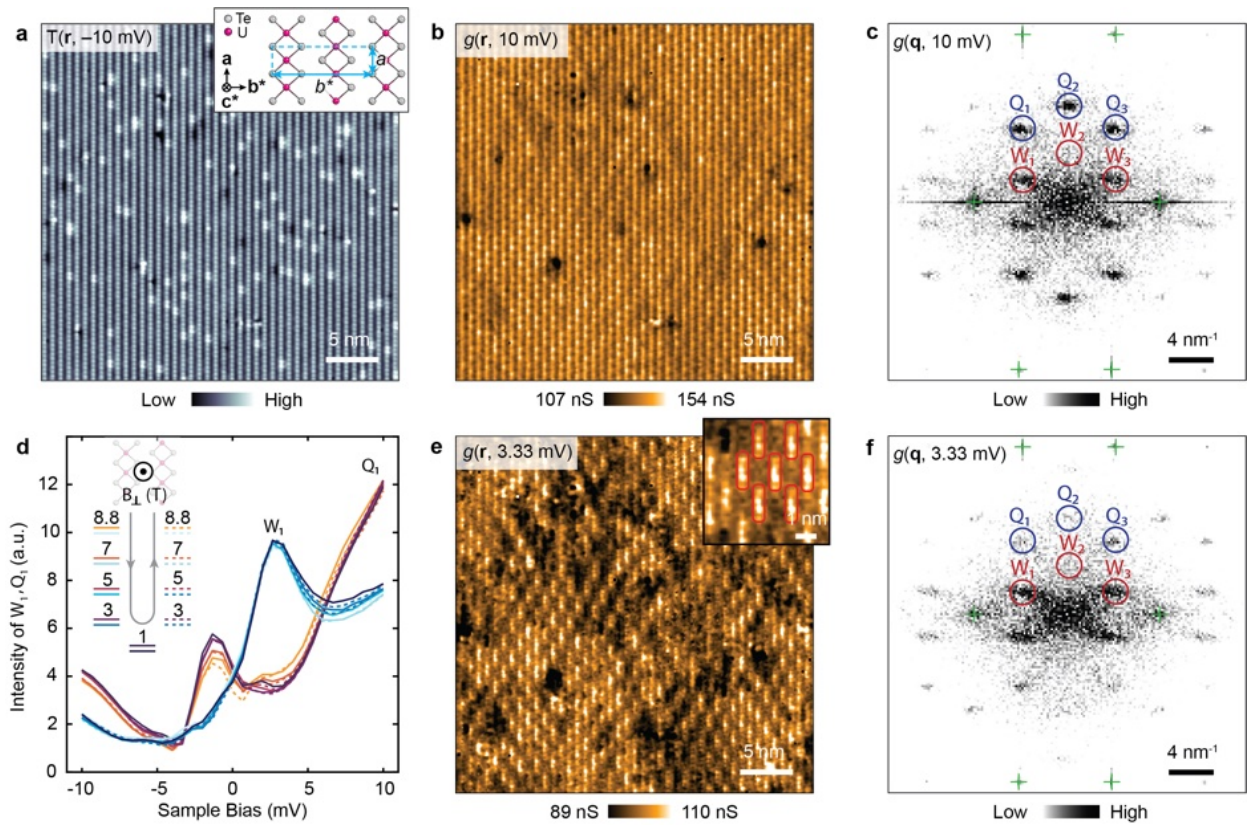

**Figure 1. Discovery of a new staggered CDW**
**a**, STM topographic image of UTe$_2$ (011) surface ($V_b$= –10 mV, $I_t$ = 1 nA). Inset: schematic of the UTe$_2$ (011) surface structure. **b**, Differential conductance image $g(\mathbf{r}, 10\text{ mV})$ in the same field of view as in (a) and **c**, its Fourier transform, acquired at 0.3 K. The Bragg peaks are indicated by green crosses, and the CDWs wavevectors are indicated by blue ($Q_i$) and red ($W_i$) circles. At this bias, the $Q_i$ CDW has maximal intensity. **d**, Bias-dependence of the CDW intensity at different $B_\perp$ and field-sweeping directions. **e**, Differential conductance image $g(\mathbf{r}, 3.33\text{ mV})$ in the same field of view as in (a) and **f**, its Fourier transform. The Bragg peaks are indicated by green crosses, and the CDWs wavevectors are indicated by blue ($Q_i$) and red ($W_i$) circles.

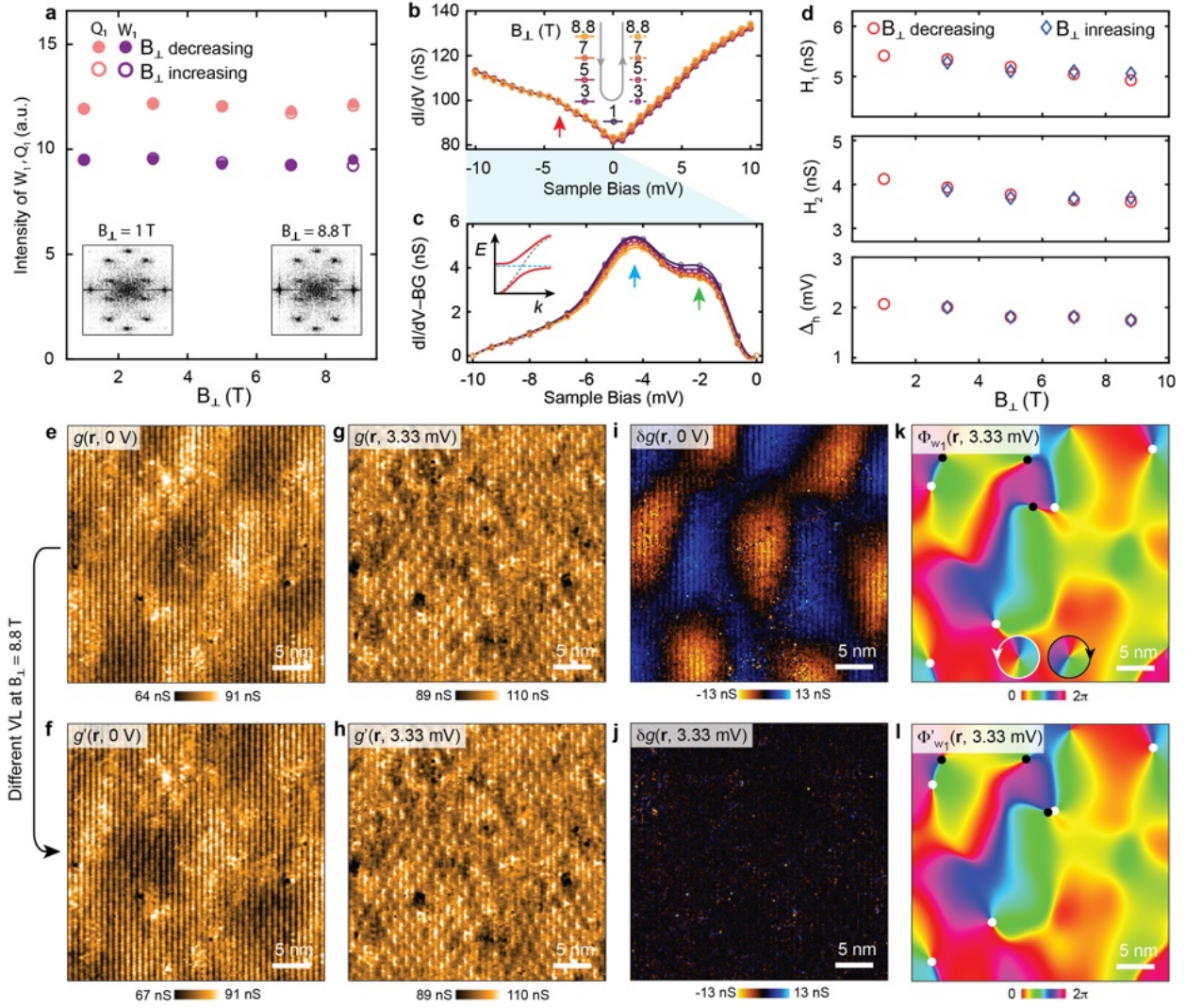

**Figure 2. Insensitivity of CDW and Kondo resonance to $B_\perp$**

**a**, Nearly unchanged extracted intensity of $W_1$ and $Q_1$ as a function of $B_\perp$. Inset: $g(\mathbf{q}, 10\text{ mV})$ images showing minimal differences between $B_\perp = 1\text{ T}$ and $8.8\text{ T}$. **b**, A series of $dI/dV$ spectra taken at 0.3 K under various $B_\perp$ and field ramping directions. **c**, Background-subtracted spectra (see Methods) in the negative energy range, showing the presence of two peaks indicated by the blue and green arrows. Inset: schematic of the hybridization between a light conduction band (grey) and a heavy $f$-electron band (blue) in a Kondo lattice. **d**, Extracted peak heights $H_{1,2}$, corresponding to the peaks indicated by the blue and green arrows in (c), respectively, and the hybridization gap $\Delta_h$, defined as the energy separation of the two peaks, as a function of $B_\perp$ and field ramping directions. **e,f**, Differential conductance images of the same field of view after two separate field cooling procedures (labeled as $g(\mathbf{q}, 0\text{ V})$ and $g'(\mathbf{q}, 0\text{ V})$) under $B_\perp = 8.8\text{ T}$, showing two displaced triangular VLs. **g,h**, Corresponding $g(\mathbf{q}, 3.33\text{ mV})$ and $g'(\mathbf{q}, 3.33\text{ mV})$ images showing virtually no differences, acquired simultaneously and in the same field of view as (e,f). **i,h**, Difference images $\delta g(\boldsymbol{r}, V) \equiv g(\boldsymbol{r}, V) - g'(\boldsymbol{r}, V)$ calculated at $V = 0$ and 3.33 mV, respectively. **k,l,** Spatial phases $\Phi_{W_1}$ and $\Phi'_{W_1}$ of the $W_1$ CDW extracted in the same field view as (e,f) with topological defects in the form vortices and anti-vortices indicated by white and black dots, respectively. A pair of such defects is shown in the inset of (k).

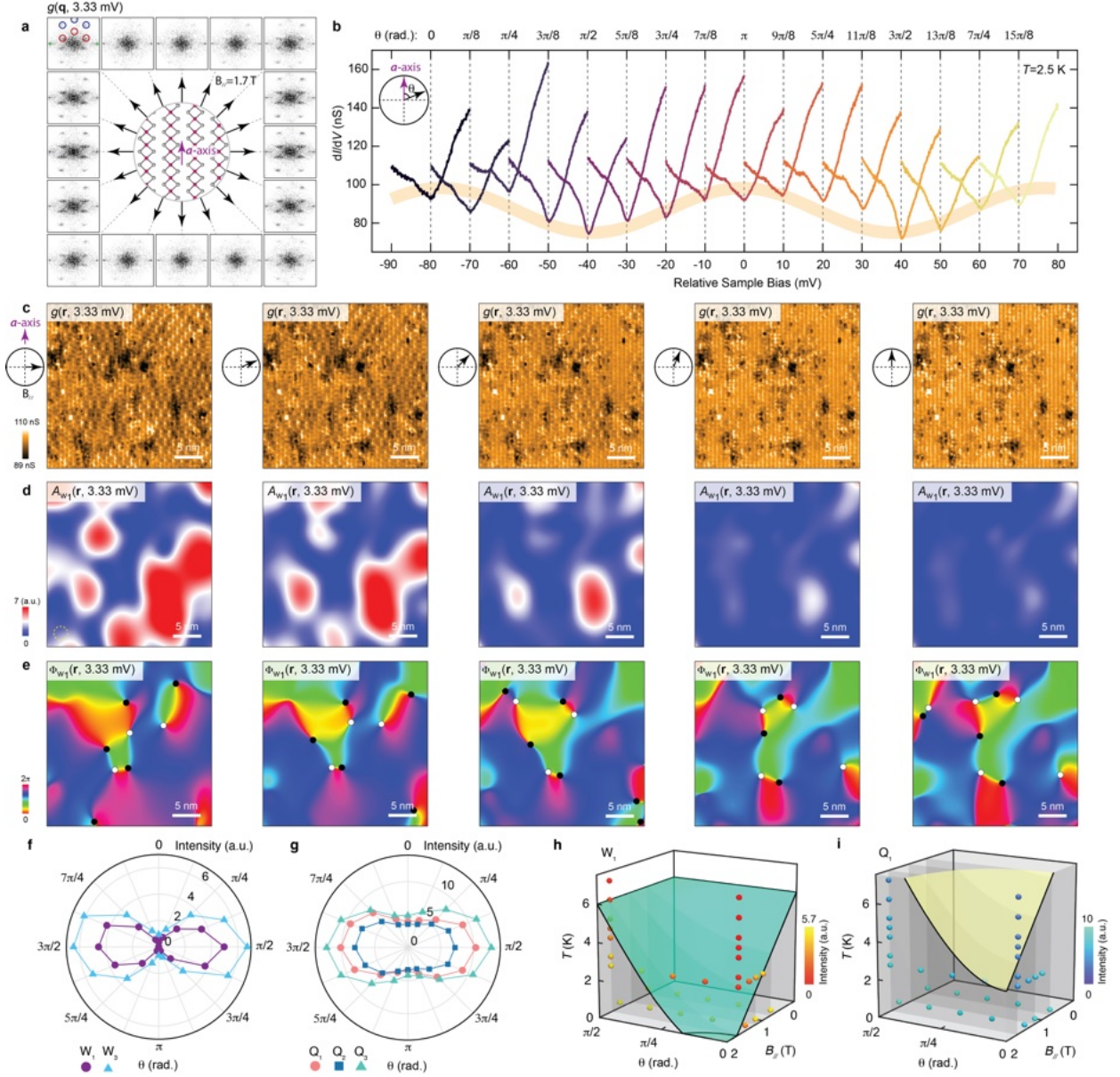

**Figure 3 Vector magnetic field manipulation of CDW**

**a**, A series of $g(\mathbf{q}, 3.33\,\text{mV})$ images acquired at 2.5 K and under a fixed $B_\perp = 1$ T, which has virtually no effect on the CDW, and $B_\parallel = 1.7$ T applied along different in-plane directions as in indicated by the black arrows. The $W_i$ and $Q_i$ CDW peaks are indicated by the red and blue circles, respectively. **b**, A series of $dI/dV$ spectra taken when the angle $\theta$ between $\boldsymbol{B}_\parallel$ and the $a$-axis of UTe$_2$ is varied between 0 and $2\pi$. The periodic modulation of the zero bias conductance is indicated by the thick orange curve. **c**, Real space images of $g(\mathbf{r}, 3.33\,\text{mV})$, **d**, the amplitude of $W_1$ modulation $A_{W_1}(\mathbf{r}, 3.33\,\text{mV})$, and **e**, the spatial phase of $W_1$ modulation $\Phi_{W_1}(\mathbf{r}, 3.33\,\text{mV})$ for $\theta = \frac{\pi}{2}, \frac{3\pi}{8}, \frac{\pi}{4}$, and $\frac{\pi}{8}$. The topological defects in $\Phi_{W_1}$ are indicated by the white and black dots. The real space Gaussian window size (see Methods) is indicated by the dashed circle in the left panel. **f**, Extracted integrated intensity of the $W_i$ and **g**, $Q_i$ CDW peaks as a function of $\theta$ from $g(\mathbf{q}, 3.33\text{ mV})$ and $g(\mathbf{q}, 10\text{ mV})$, respectively, showing clear in-plane anisotropy. **h**, Color-coded intensity of $W_1$ and **i**, $Q_1$ CDWs plotted in the phase space of $B_\parallel$, $\theta$, and $T$. The fitted phase boundaries defining the CDW states are colored green and yellow, respectively.

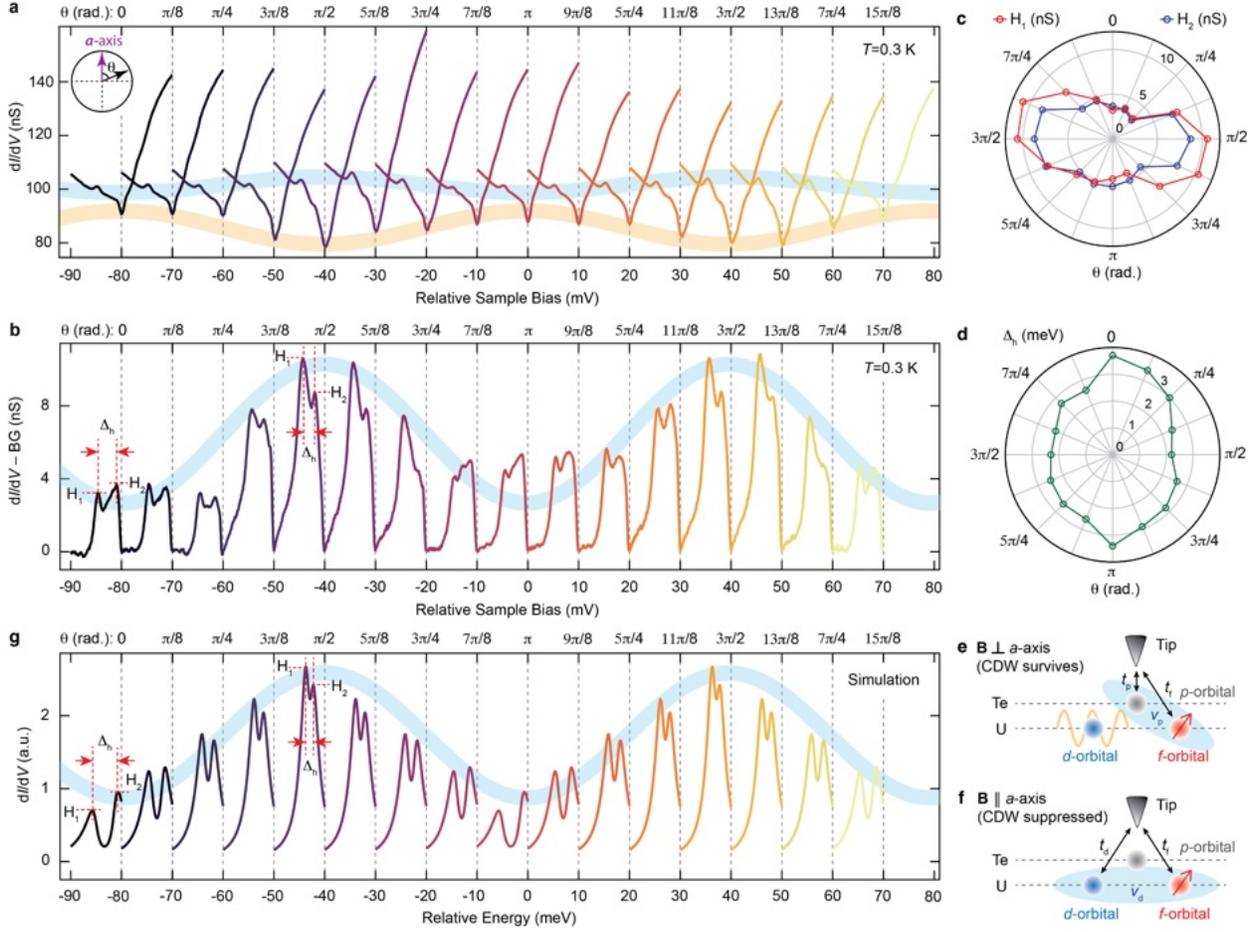

**Figure 4 Vector magnetic field manipulation of Kondo effect**
**a,** A series of d$I$/d$V$ spectra taken at 0.3 K under $B_{\parallel} = 1.95$ T when the angle $\theta$ between $\boldsymbol{B}_{\parallel}$ and the $a$-axis of UTe$_2$ is varied between 0 and $2\pi$. The periodic modulation of the zero bias conductance is indicated by the thick orange curve. An anti-correlated modulation of the Kondo resonance feature is indicated by the thick blue curve. **b,** Corresponding background-subtracted spectra of (a), highlighting the modulation of the Kondo resonance. The peak heights ($H_{1,2}$) and hybridization gap ($\Delta_h$) are indicated at $\theta = 0$ and $\pi/2$. **c,** Extracted angular dependence of $H_{1,2}$ and **d,** $\Delta_h$, showing clear anisotropy and anti-correlation. **e,** When the in-plane field $\boldsymbol{B}_{\parallel} \perp a$-axis ($\theta = \pi/2$), the U-6$d$ orbital states develop CDW order and the effective Kondo hybridization (blue oval) is in the $p$-$f$ channel and probed by an STM tip. **f,** When $\boldsymbol{B}_{\parallel} \parallel a$-axis ($\theta = 0$), the CDW order in U-6$d$ orbital states is suppressed and the effective Kondo hybridization is in the $d$-$f$ channel and probed by an STM tip. Here, $t_p$, $t_d$ and $t_f$ are the tunneling strengths between the STM tip to the Te-5$p$, U-6$d$, and U-5$f$ electrons, respectively; $v_p$ and $v_d$ are the hybridization amplitudes of U-5$f$ with Te-5$p$ and U-6$d$, respectively. **g,** A series of calculated d$I$/d$V$ spectra as $\theta$ evolves, reproducing qualitatively the experimental

d$I$/d$V$ spectra in (b). The calculation details and tight-binding model parameters of U-6$d$ and Te-5$p$ orbitals are given in the Methods section.

# Directional Manipulation of a Staggered Charge Density Wave and Kondo Resonance in UTe$_2$

Nileema Sharma[1,2][†], Fangjun Cheng[1,2][†], Hyeok Jun Yang[1][†], Matthew Toole[1,2], James McKenzie[1,2], Mitchell M. Bordelon[3], Sean M. Thomas[3], Priscila F. S. Rosa[3], Yi-Ting Hsu[1]*, and Xiaolong Liu[1,2]*

[1]*Department of Physics and Astronomy, University of Notre Dame, Notre Dame, IN 46556, USA*

[2]*Stavropoulos Center for Complex Quantum Matter, University of Notre Dame, Notre Dame, IN 46556, USA*

[3]*Los Alamos National Laboratory, Los Alamos, New Mexico 87545, USA*

* Corresponding authors. yhsu2@nd.edu, xliu33@nd.edu

† These authors contributed equally to this work.

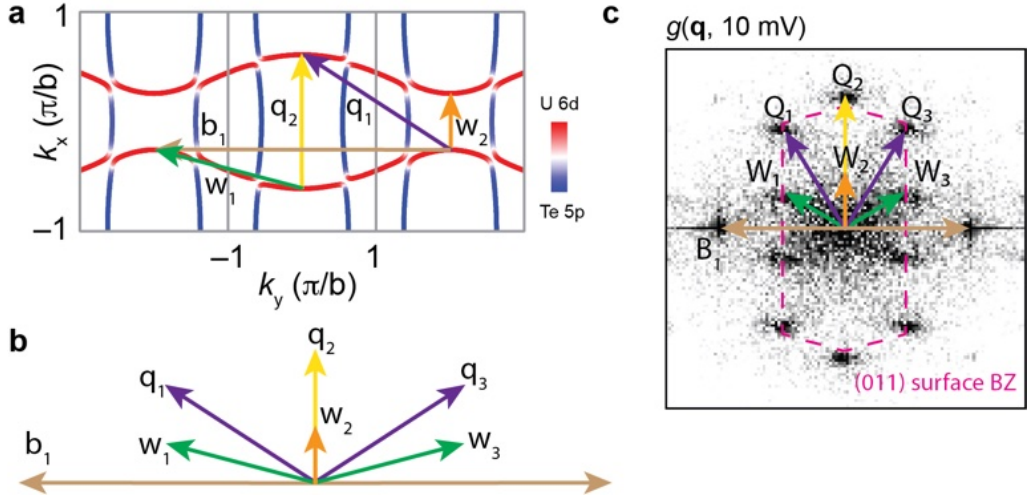

**Supplementary Figure 1. Fermi surface and nesting vectors of UTe$_2$**
**a.** Fermi surface (FS) of UTe$_2$ in the $k_y$ (horizontal axis) $-$ $k_x$ (vertical axis) plane at $k_z = 0$, color coded to reflect the orbital contributions. The quasi-1D FSs primarily composed of U 6d and Te 5p orbitals run along $k_y$ and $k_x$ directions, respectively, and hybridize at their intersections. As identified in Ref. 32, the nesting vectors connect different portions of the FS involving only the U 6d orbitals (red). Among them, the $b_1$ vector is a Bragg vector. **b**, The same nesting vectors as in (a), but plotted with the origins together. **c**, Experimental $g(\boldsymbol{q}, 10 \text{ mV})$ image on the UTe$_2$ (011) surface with overlaid nesting vectors transformed to the (011) plane. To do so, we follow the same procedure demonstrated in ref. 32, where the transformation to the (011) plane corresponds to a change of $y$-axis coordinates as $\boldsymbol{Q}_y = \boldsymbol{q}_y \sin\theta$, where $\theta = 23.7°$ is angle between the (011) surface normal and the $b$-axis. The resulting nesting wavevectors match closely the experimental ones. The (011) surface Brillouin zone (BZ) is shown in pink, constructed from the six Bragg vectors in Fig. 1c (green crosses).

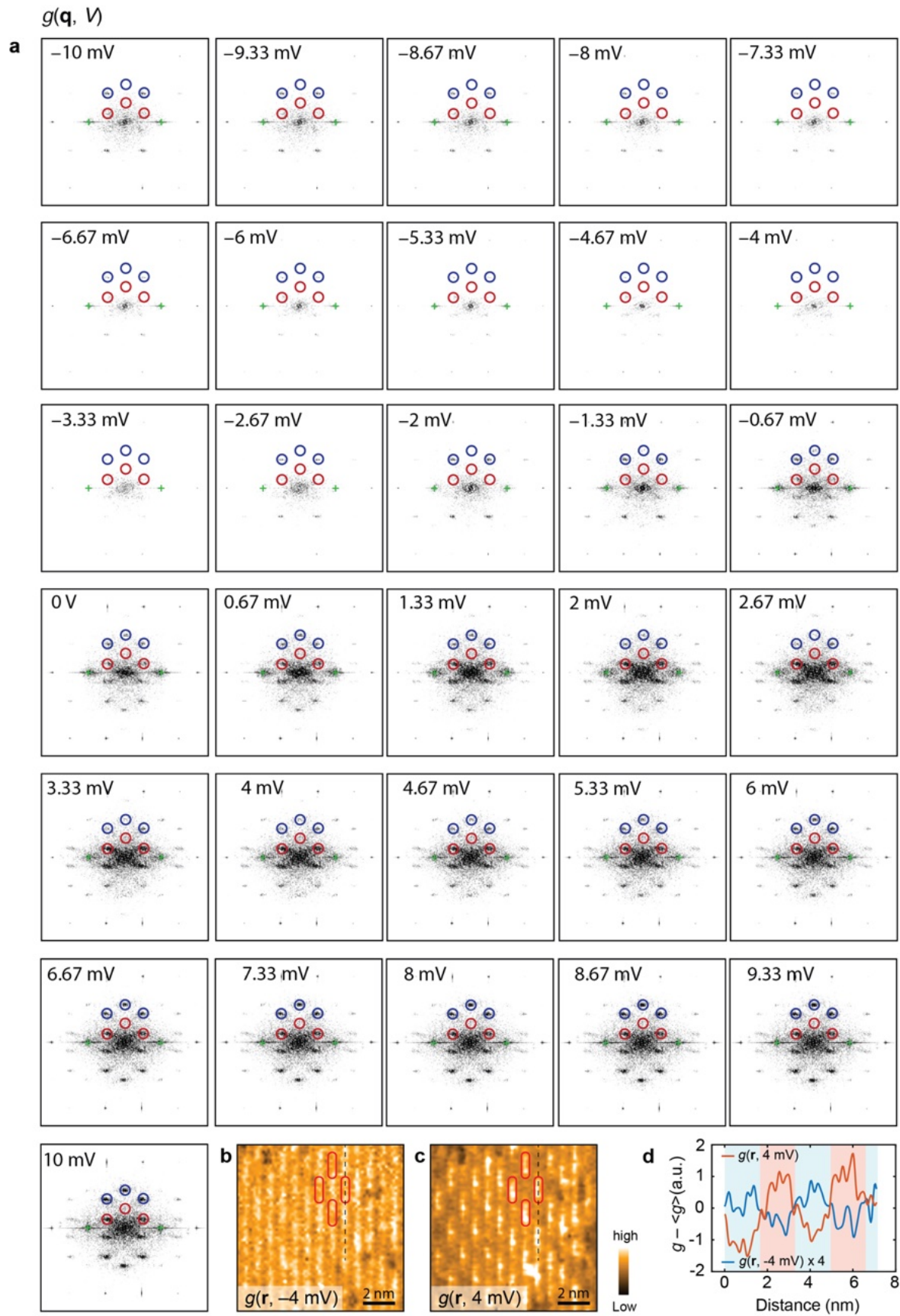

**Supplementary Figure 2. Bias-dependence of CDWs and contrast inversion**
**a,** A series of $g(\boldsymbol{q},V)$ images in both the filled ($V<0$) and empty states ($V>0$) at 0.3 K. The CDWs are indicated by circles. **b**, $g(\boldsymbol{r},-4\text{ mV})$ and **c**, $g(\boldsymbol{r},4\text{ mV})$ images in the same FOV, showing both the brick-wall-like pattern from the staggered $W_i$ CDW, but with a contrast inversion. To better visualize this contrast inversion, we highlight four "bricks" in red. **d**, Line profiles of $g(\boldsymbol{r},V)-<g(\boldsymbol{r},V)>$ along the dashed lines in (b,c), where $<g(\boldsymbol{r},V)>$ is the averaged differential conductance of the entire FOV. Contrast inversion in the filled and empty states are clearly seen, consistent with a CDW from Fermi surface nesting.

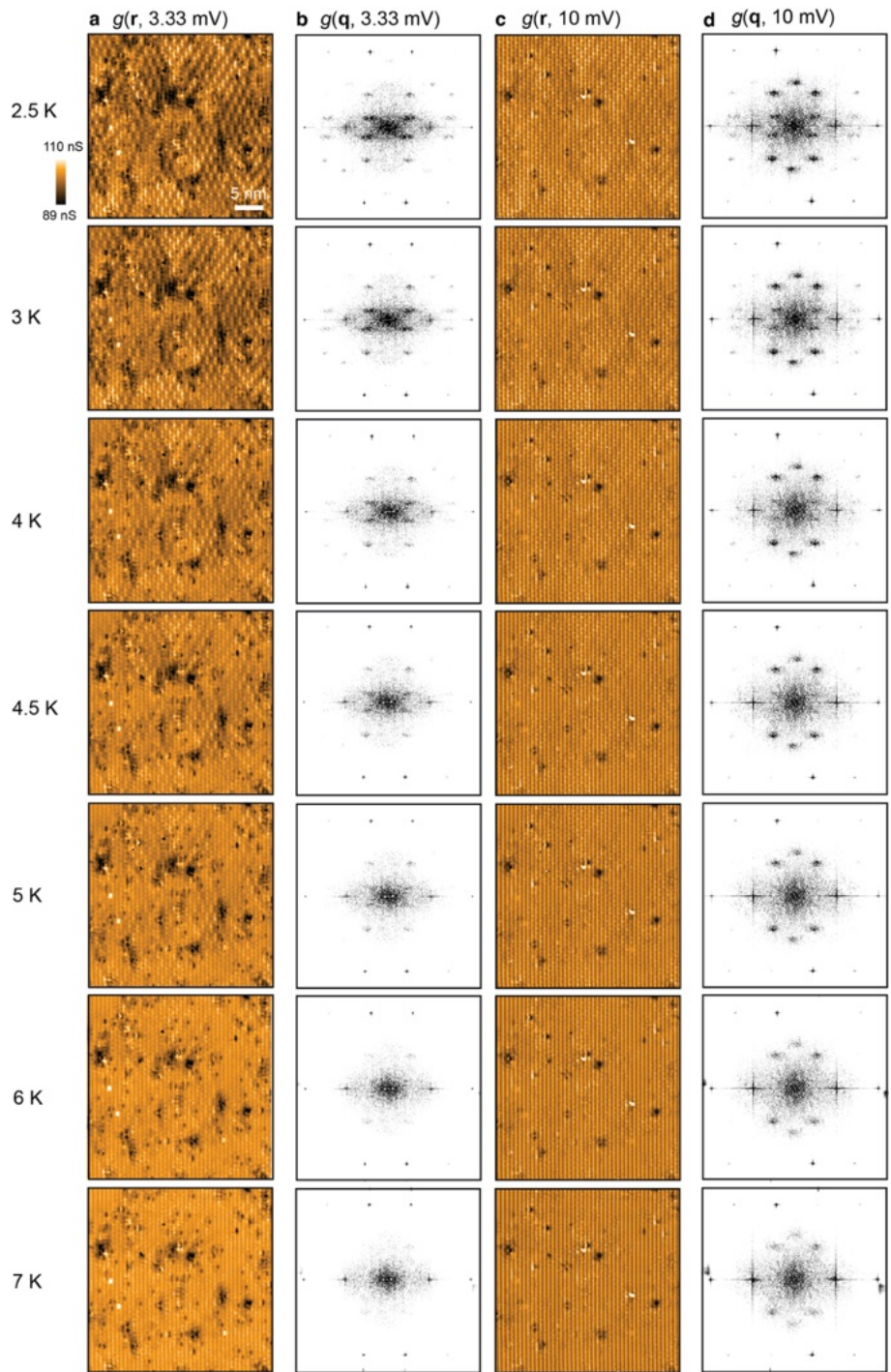

**Supplementary Figure 3. Thermal melting of CDWs**
**a,** A series of $g(r, 3.33 \text{ mV})$, **b,** $g(q, 3.33 \text{ mV})$, **c,** $g(r, 10 \text{ mV})$, and **d,** $g(q, 10 \text{ mV})$ images acquired in the same FOV at different temperatures, showing clearly the thermal melting of the $W_i$ and $Q_i$ CDWs. **e,** Intensity of the $W_i$ and **f,** $Q_i$ CDW (extracted from the FTs) as a function of temperature.

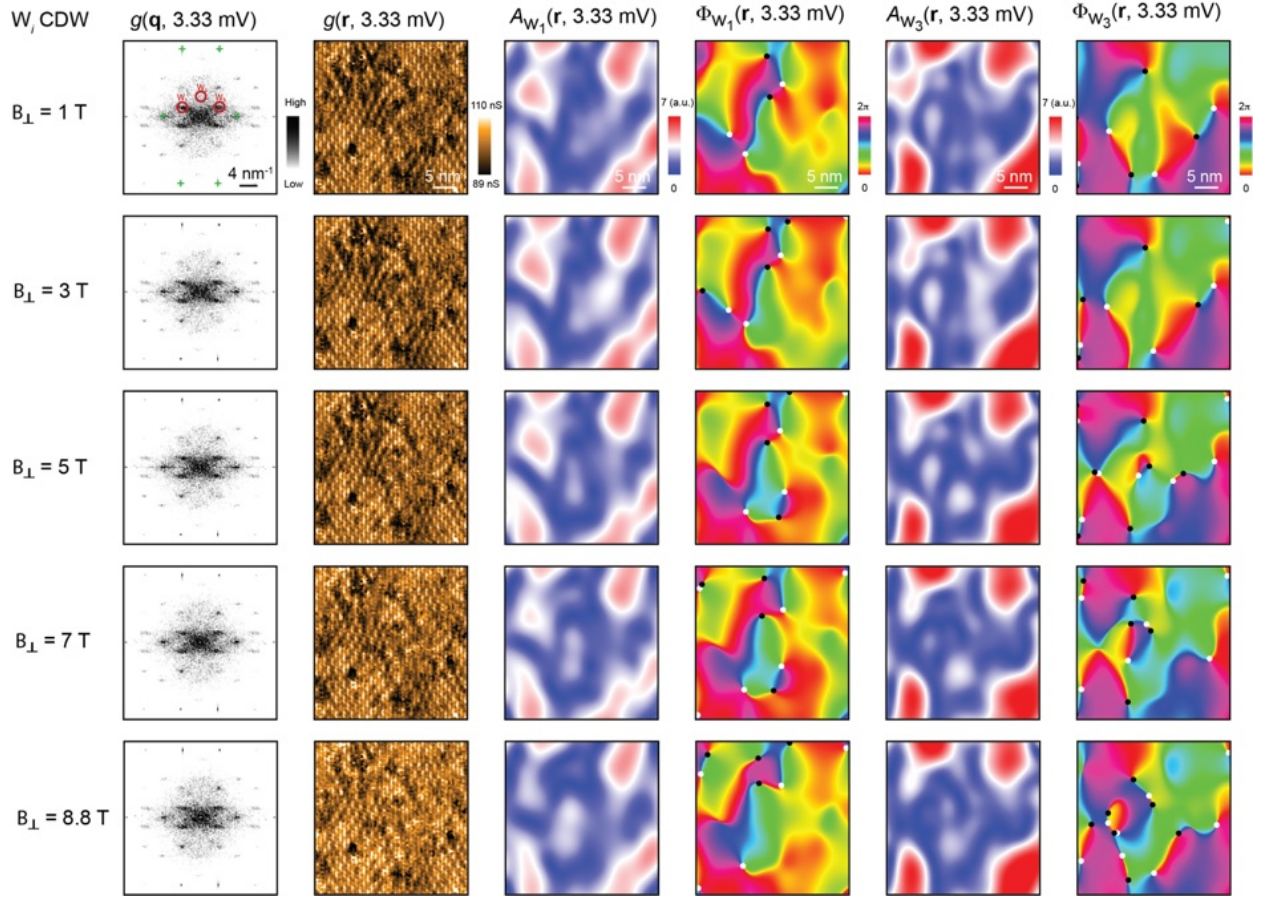

**Supplementary Figure 4. Evolution of $W_i$ CDW as a function of $B_\perp$**
In the same FOV, the momentum-space ($g(\mathbf{q}, 3.33\,\text{mV})$) and real-space ($g(\mathbf{r}, 3.33\,\text{mV})$) structures of $W_i$ CDW are visualized under different out-of-plane field $B_\perp$ at 0.3 K, showing minimal changes. Such insensitivity of the $W_i$ CDW to $B_\perp$ is further visualized by extracting the CDW amplitude ($A_{W_i}(\mathbf{r}, 3.33\,\text{mV})$, see Methods) and spatial phase ($\Phi_{W_i}(\mathbf{r}, 3.33\,\text{mV})$, see Methods) images, which also do not evolve with $B_\perp$ in a systematic manner. The topological defects in $\Phi_{W_i}$ are indicated by the white and black dots. Real space correspondence between the mirror symmetric $W_1$ and $W_3$ CDW is observed clearly from their similar amplitude ($A_{W_i}$) distributions.
7

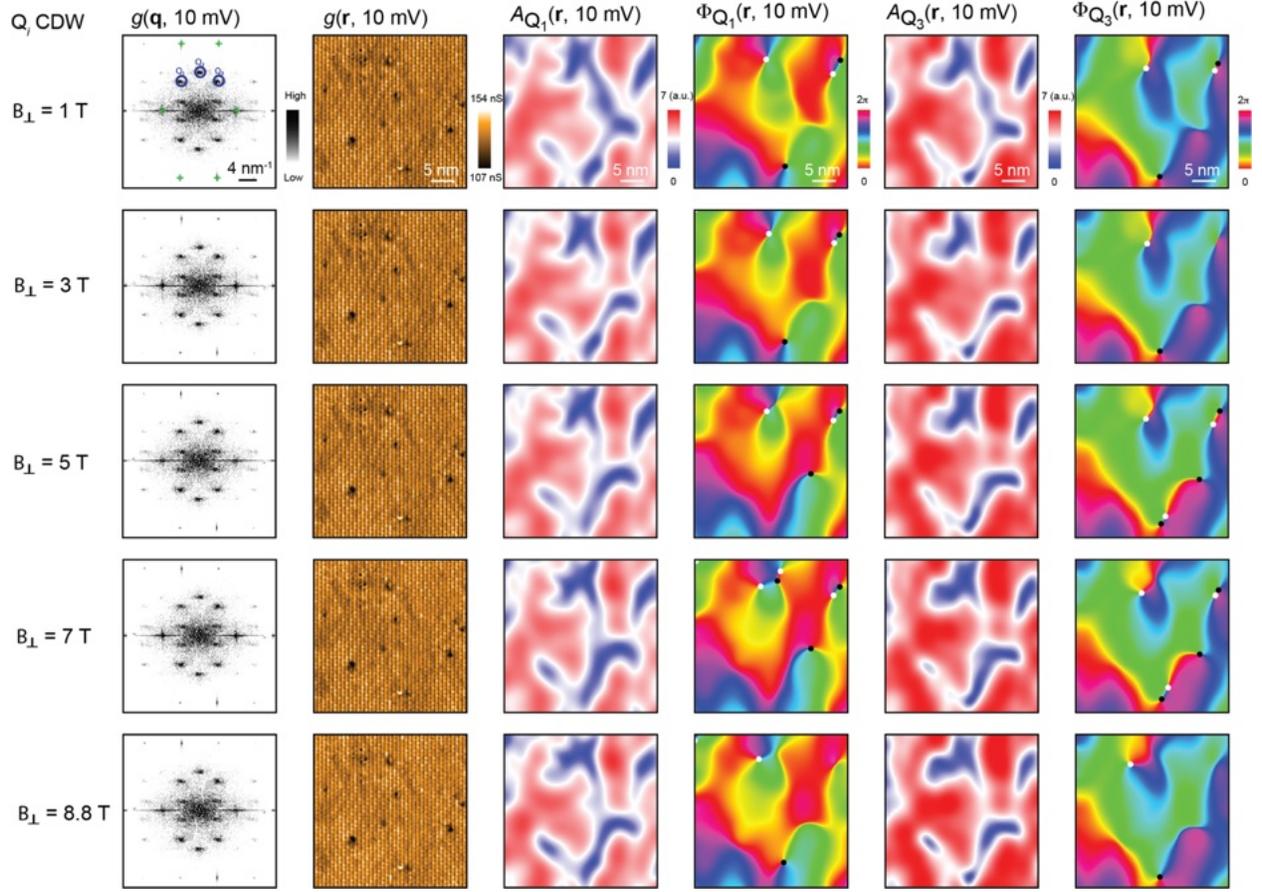

**Supplementary Figure 5. Evolution of $Q_i$ CDW as a function of $B_\perp$**
In the same FOV, the momentum-space ($g(\mathbf{q}, 10\text{ mV})$) and real-space ($g(\mathbf{r}, 10\text{ mV})$) structures of $Q_i$ CDW are visualized under different out-of-plane field $B_\perp$ at 0.3 K, showing minimal changes. Such insensitivity of the $Q_i$ CDW to $B_\perp$ is further visualized by extracting the CDW amplitude ($A_{Q_i}(\mathbf{r}, 3.33\text{ mV})$, see Methods) and spatial phase ($\Phi_{Q_i}(\mathbf{r}, 3.33\text{ mV})$, see Methods) images, which also do not evolve with $B_\perp$ in a systematic manner. The topological defects in $\Phi_{Q_i}$ are indicated by the white and black dots. Real space correspondence between the mirror symmetric $Q_1$ and $Q_3$ CDW is observed clearly from their similar amplitude ($A_{Q_i}$) distributions.

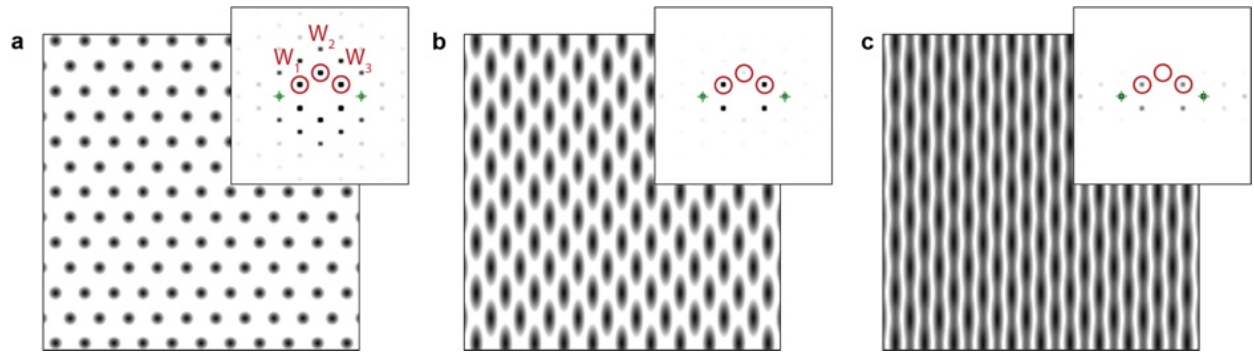

**Supplementary Figure 6. Effect of CDW formfactor on the Fourier transform**
**a-c,** Simulated triangular CDW lattices with increasing deviations from an isotropic formfactor, mimicking the staggered brick-wall pattern of $W_i$ CDW. The FTs shown as insets clearly demonstrate the diminishing peak correspond to $W_2$ CDW. Therefore, the formfactor of the $W_i$ CDW might have contributed to the much lower experimental intensity of $W_2$.

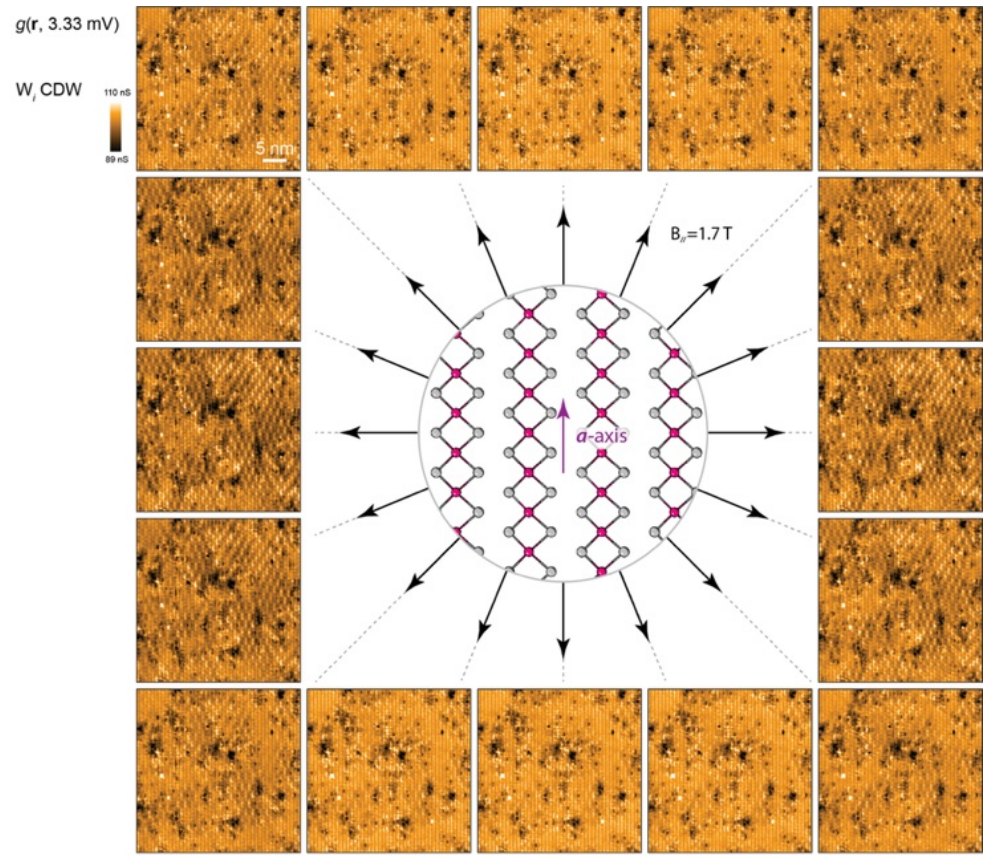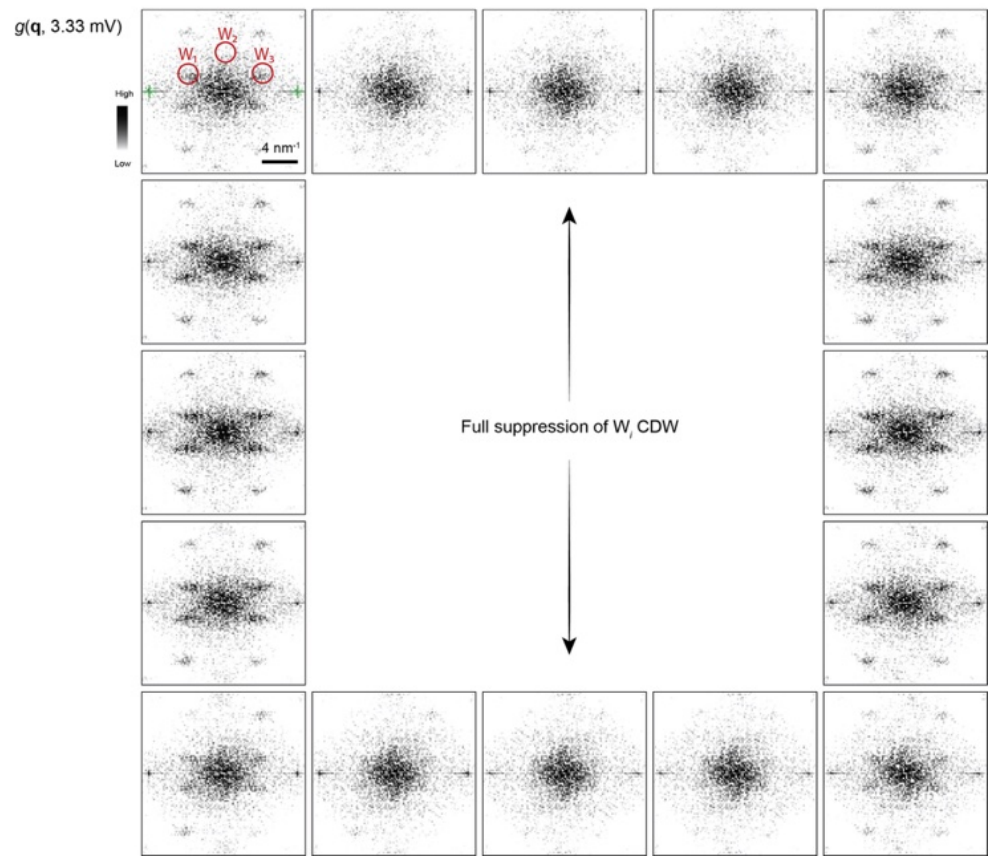

**Supplementary Figure 7. In-plane field manipulation of $W_i$ CDW**

A series of $g(\mathbf{r}, 3.33 \text{ mV})$ (top) and $g(\mathbf{q}, 3.33 \text{ mV})$ (bottom) images of the same FOV but under different in-plane field (1.7 T) directions measured at 2.5 K. Complete suppression of $W_i$ (indicated by red circles in the FTs) is observed when the field is aligned (or anti-aligned) with the $a$-axis of UTe$_2$.

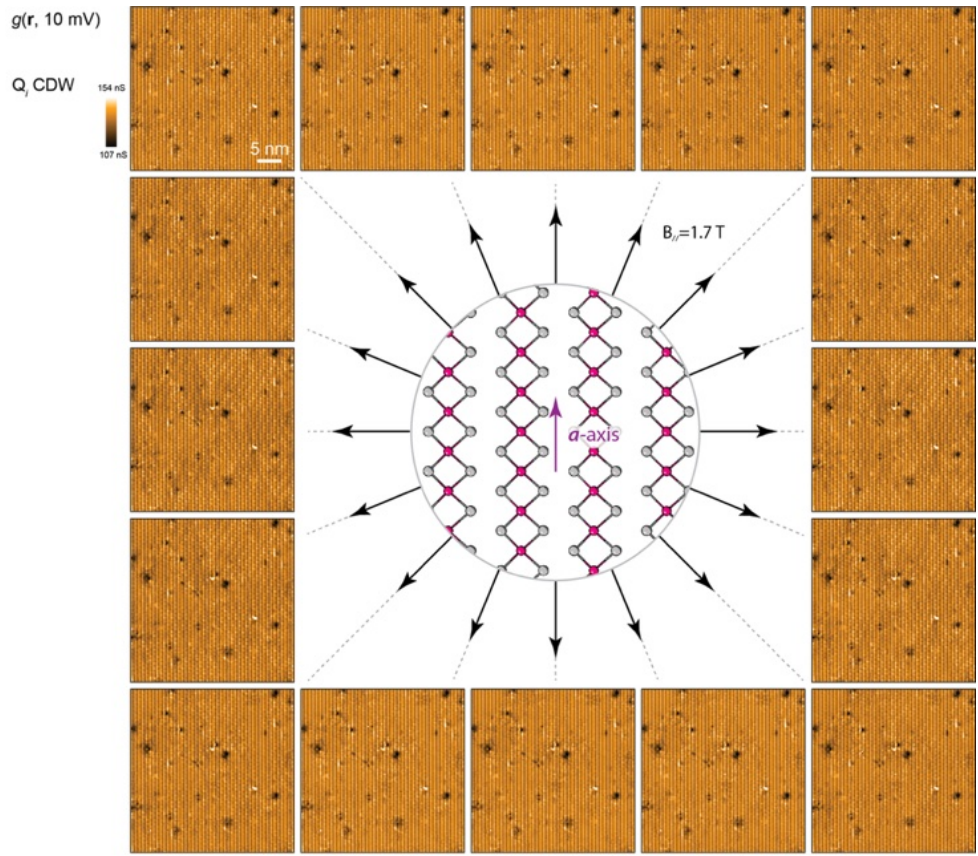
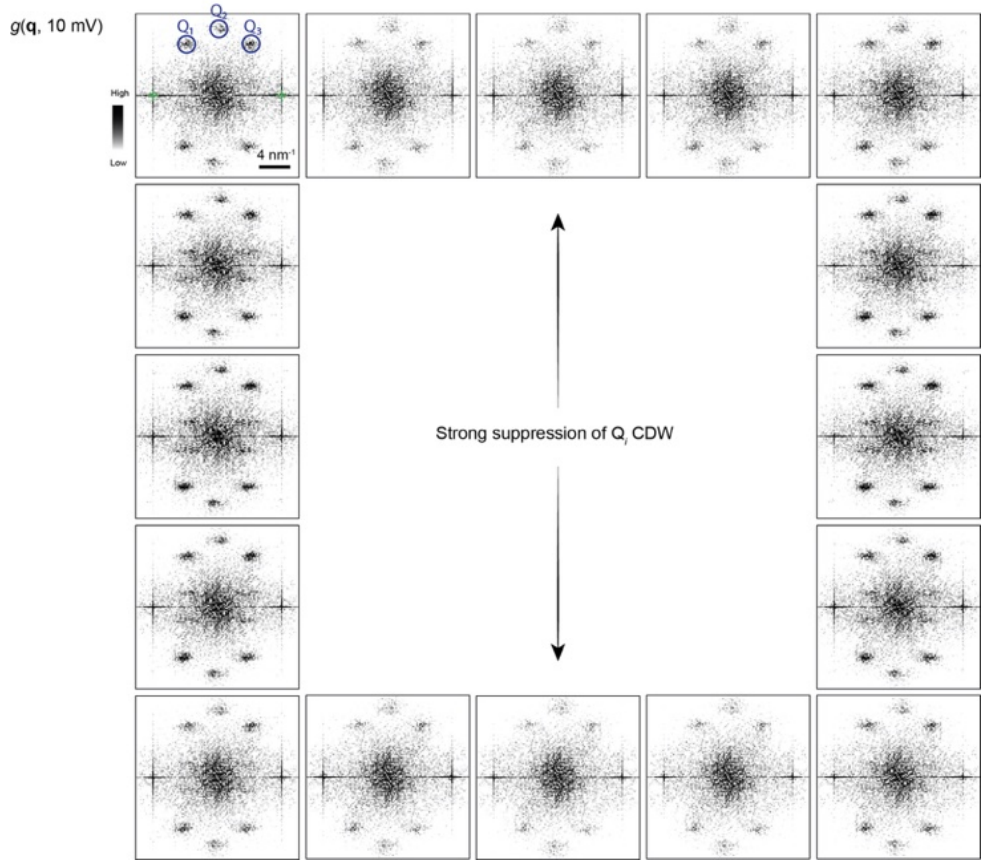

**Supplementary Figure 8. In-plane field manipulation of $Q_i$ CDW**
A series of $g(\boldsymbol{r}, 10\text{ mV})$ (top) and $g(\boldsymbol{q}, 10\text{ mV})$ (bottom) images of the same FOV but under different in-plane field (1.7 T) directions measured at 2.5 K. Strong but not complete suppression of $Q_i$ (indicated by blue circles in the FTs) is observed when the field is aligned (or anti-aligned) with the *a*-axis of UTe$_2$.

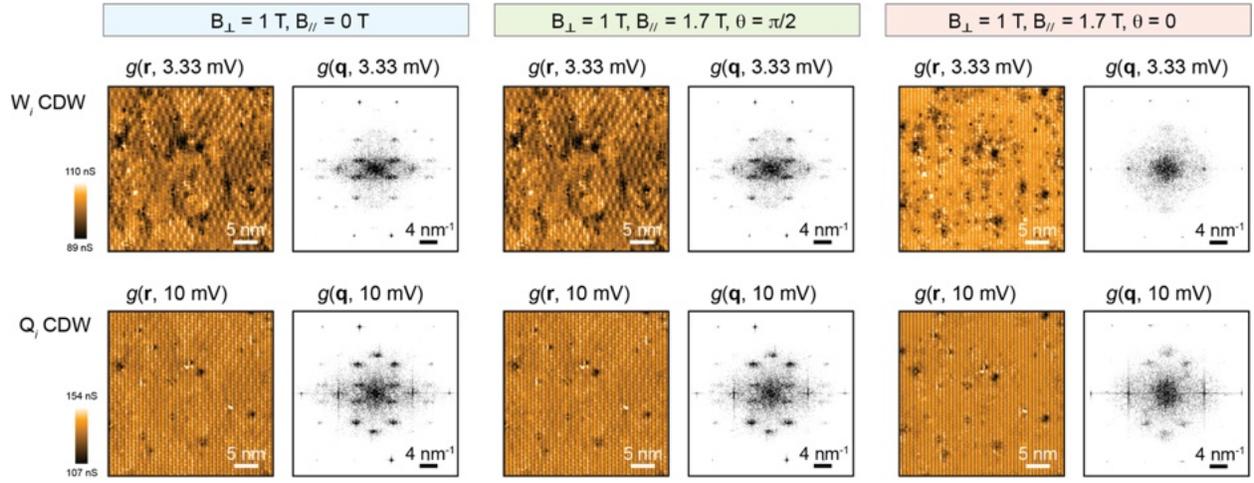

**Supplementary Figure 9. Minimal influence of in-plane field along the $b^*$-axis**
A series of $g(r, 3.33\text{ mV})$ (characterizing $W_i$ CDW), $g(r, 10\text{ mV})$ (characterizing $Q_i$ CDW), and their FTs are presented under different in-plane field configurations at 2.5 K: (i) $B_\parallel = 0$, (ii) $B_\parallel = 1.7$ T and $\theta = \pi/2$ (i.e., along the $b^*$-axis, Fig. 1a), and (iii) $B_\parallel = 1.7$ T and $\theta = 0$ (i.e., along the $a$-axis, Fig. 1a). While virtually no difference can be observed between cases (i) and (ii), demonstrating the minimal influence of in-plane field of 1.7 T along the $b^*$-axis, a complete (strong) suppression of the $W_i$ and $Q_i$ CDW takes place when $B_\parallel$ is along the $a$-axis direction.

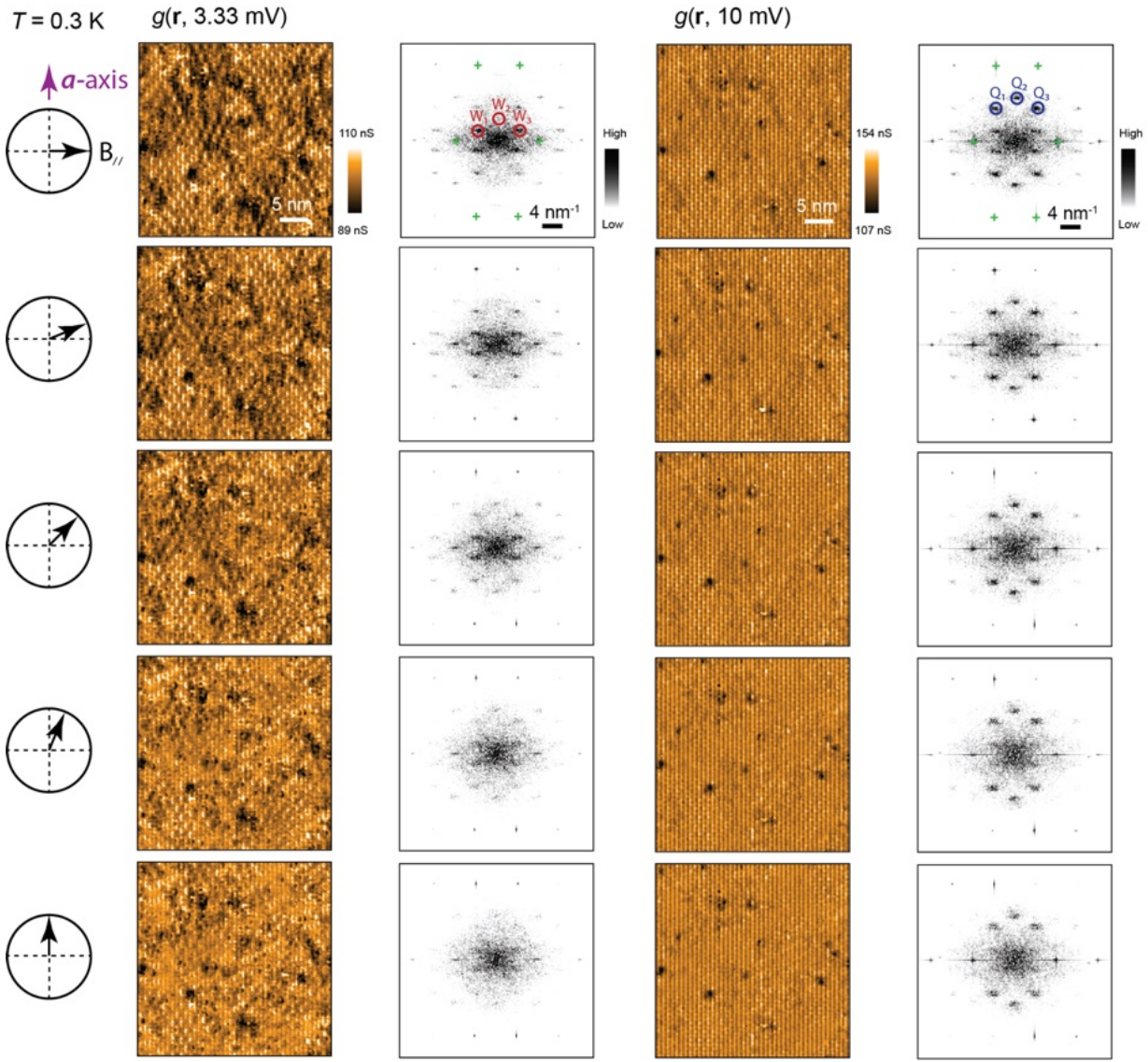

**Supplementary Figure 10. Strong directional field sensitivity of CDWs at 0.3 K**
A series of $g(\mathbf{r}, 3.33\ \text{mV})$ (characterizing $W_i$ CDW), $g(\mathbf{r}, 10\ \text{mV})$ (characterizing $Q_i$ CDW), and their FTs are presented under different in-plane field $B_\parallel = 1.5\ \text{T}$ directions with $\theta$ varying between 0 and $\pi/2$, which is indicated schematically. Similar to the observations at 2.5 K (Fig. 3, Supplementary Figs. 7,8), a complete (strong) suppression of the $W_i$ and $Q_i$ CDW takes place when $B_\parallel$ is along the $a$-axis direction ($\theta = 0$).

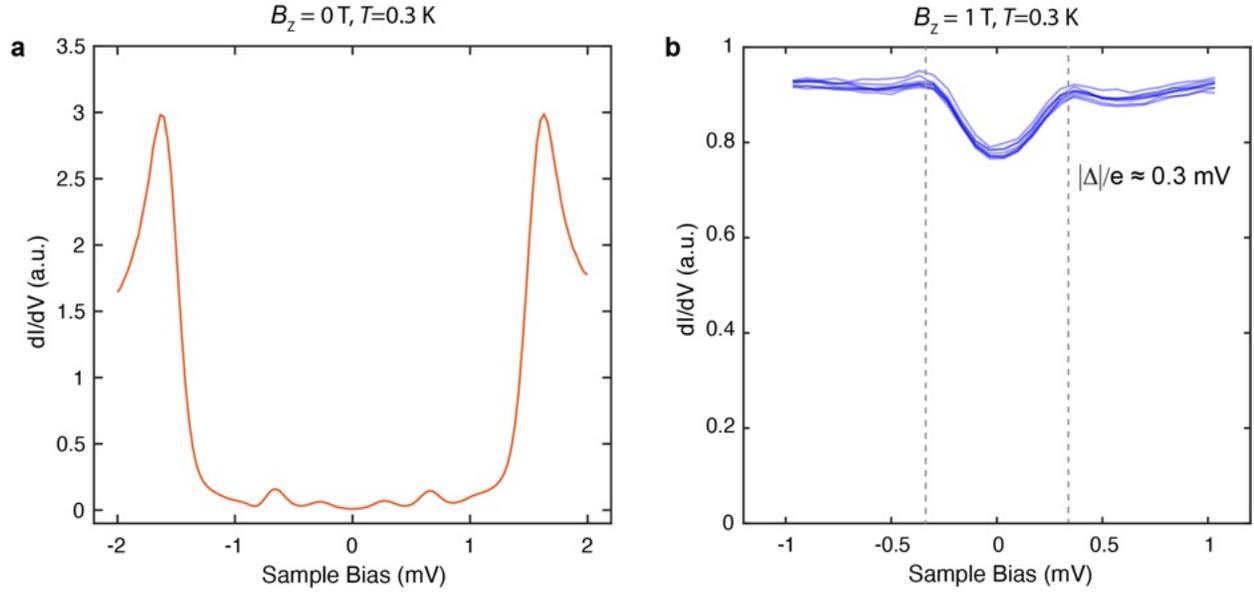

**Supplementary Figure 11. Tunneling spectroscopy between the Nb tip and UTe₂**
**a,** Representative d$I$/d$V$ spectrum of UTe$_2$ measured with a Nb tip at 0.3 K and zero external magnetic field ($B = 0$ T), showing the total coherence peaks and (multiple) Andreev reflection peaks as expected in a superconducting-vacuum-superconducting junction. **b**, d$I$/d$V$ spectra of UTe$_2$ measured with a Nb tip at 0.3 K at different locations on the sample under an external field of $B_z = 1$ T (away from any vortices). Such an external field suppresses tip superconductivity and makes the tip normal but has minimal influence on UTe$_2$ superconductivity given its exceedingly high critical fields, thereby showing characteristic coherence peaks of high-quality UTe$_2$ at $\pm\frac{|\Delta|}{e} \approx$ 0.3 mV. In this study, the applied field in any direction exceeds 1 T, and therefore the tip remains normal in all measurements.